\documentclass[12pt,a4paper]{article}

\usepackage[utf8]{inputenc}
\usepackage[T1]{fontenc}
\usepackage{amsmath,amssymb,amsthm,amsfonts}
\usepackage{bm}
\usepackage{graphicx}
\usepackage{booktabs}
\usepackage{geometry}
\usepackage{hyperref}
\usepackage{natbib}
\usepackage{mathptmx}
\usepackage{authblk}
\usepackage{float}
\usepackage{pifont}
\hypersetup{
	colorlinks=true,
	linkcolor=blue,
	citecolor=blue,
	urlcolor=blue
}

\newcommand{\ind}{\mathbb{I}}
\newcommand{\bY}{\bm{Y}}
\newcommand{\bS}{\bm{S}}
\newcommand{\bTheta}{\bm{\Theta}}
\newcommand{\bpi}{\bm{\pi}}
\newcommand{\bGamma}{\bm{\Gamma}}
\newcommand{\bgamma}{\bm{\gamma}}
\newcommand{\btheta}{\bm{\theta}}

\newcommand{\bepsilon}{\bm{\epsilon}}

\newcommand{\E}{\mathbb{E}}

\newcommand{\R}{\mathbb{R}}
\newcommand{\N}{\mathbb{N}}

\theoremstyle{plain}
\newtheorem{theorem}{Theorem}[section]

\theoremstyle{definition}

\newtheorem{remark}[theorem]{Remark}

\title{
	\bfseries
	A Bayesian Hidden Markov Model with \\
	Singh--Maddala Emission Distributions for \\
	Heavy-Tailed Regime-Switching Time Series
}

\author[1,2]{Safaa K. Kadhem \thanks{Corresponding author: safaa.kadhem@mu.edu.iq}}
\affil[1]{Department of Mathematics and Computer Applications, College of Science, Al-Muthanna University, Samawah, Iraq}
\affil[2]{Department of Artificial Intelligence Engineering, College of Engineering, Al-Alyen Iraqi University, Thi-Qar, Iraq}

\date{\today}

\begin{document}
	\maketitle
	
	% ============================================================
	% ABSTRACT — بعد الحذف: فقرتان
	% ============================================================
	\begin{abstract}
		\noindent
		
		Heavy-tailed time series that exhibit regime switching are
		common  in environmental, financial, and engineering
		applications, yet their statistical modeling remains
		challenging because conventional emission distributions
		--- such as the Gaussian, Gamma, or Weibull --- fail to
		capture simultaneously the extreme right-skewness, the
		slowly decaying upper tails, and the temporal dependence
		that characterize such processes.
		
		This paper develops a fully Bayesian Hidden Markov Model
		with Singh--Maddala (Burr Type~XII) emission
		distributions (HMMSM). We derive the complete Bayesian
		structure of the model --- joint prior, complete-data
		likelihood, marginal likelihood via the forward
		recursion, joint posterior, and full conditional
		distributions --- and design an adaptive
		Metropolis-within-Gibbs sampler that combines exact
		Forward-Filtering Backward-Sampling for the latent
		states, conjugate Dirichlet updates for the transition
		structure, and adaptive Metropolis steps on the
		unconstrained log-scale for the emission parameters.
		Extensive simulation studies confirm parameter recovery
		with relative errors below $10\%$ and a $95\%$
		prediction-interval coverage of $94.5\%$. On a simulated
		benchmark with heavy tails ($\alpha_3 = 2.25$), Bayesian
		model selection via WAIC, DIC, and LOO correctly
		identifies $K = 3$, and HMMSM outperforms seven competing
		specifications by large differences
		($\Delta\text{WAIC} \geq 51.76$). On a real 26-year
		gridded rainfall record for the Republic of Ireland
		($T = 312$ monthly observations, 2000--2025), HMMSM
		attains the lowest WAIC among all candidates but is
		statistically indistinguishable from HMM-Gamma
		($\Delta\text{WAIC} = 1.04$), reflecting the relatively
		light tails of Irish monthly rainfall
		($\hat{\alpha}_1 = 9.60$, $\hat{\alpha}_2 = 10.21$). The
		selected two-regime model identifies a Moderate regime
		(conditional mean $\approx 87$~mm) and a Wet regime
		($\approx 134$~mm) that are persistent and seasonally
		coherent, with the Wet regime dominating autumn and
		winter. We also document a fundamental weak
		identifiability of the Singh--Maddala scale parameter in
		this application and recommend that applied users report
		inference on the identifiable parameters as their
		primary targets.
		
		The framework is general-purpose and applicable to any
		positive-valued process with heavy tails and latent
		regime dynamics. Open-source, reproducible code is provided
		for reproducibility.
		
		\vspace{0.5cm}
		
		
	\end{abstract}
	
	\vspace{1cm}
	
	% ============================================================
	% SECTION 1: INTRODUCTION — بعد الحذف: ~2.5 صفحة
	% ============================================================
	\section{Introduction}
	\label{sec:introduction}
	
	Heavy-tailed time series that exhibit regime switching arise
	in contexts as diverse as extreme precipitation
	\citep{katz2002, wilks2011}, financial asset returns
	\citep{cont2001}, insurance claim sizes \citep{embrechts1997},
	and reliability lifetimes \citep{lawless2003}. Three features
	unite these phenomena and render their statistical modeling
	particularly challenging. First, they exhibit strong temporal
	dependence: successive observations are correlated and cluster
	in time. Second, they display pronounced right-skewness with
	heavy upper tails, so that rare but extreme events contribute
	disproportionately to overall variability. Third, they exhibit
	regime switching: the underlying data-generating mechanism
	alternates between qualitatively distinct states, such as calm
	versus crisis periods in financial markets
	\citep{hamilton1989} or dry versus stormy periods in
	atmospheric systems \citep{betro2008}.
	
	Traditional approaches address only one or two of these
	features at a time. Marginal distributions such as the Gamma,
	Weibull, and Lognormal families can capture right-skewness but
	ignore temporal dependence and regime structure
	\citep{katz2002}. Autoregressive models accommodate temporal
	dependence but typically rely on Gaussian innovations, which
	are ill-suited to heavy-tailed data \citep{cont2001}.
	Regime-switching models such as Hidden Markov Models (HMMs)
	can capture latent state dynamics but are often paired with
	emission distributions whose tails decay too rapidly to
	reproduce the extreme events that motivate the analysis in the
	first place \citep{betro2008, chida2025}. Even heavy-tailed
	alternatives such as the Gamma or Lognormal families, when
	embedded in an HMM, provide only a single tail shape and
	cannot adapt to the possibility that different regimes might
	exhibit different tail behaviors.
	
	This paper addresses the gap by developing a unified Bayesian
	framework that accommodates temporal dependence, latent regime
	structure, and heavy-tailed emissions simultaneously. The
	proposed model combines a Hidden Markov Model with
	Singh--Maddala (Burr Type~XII) emission distributions and
	implements a fully Bayesian inferential framework via Markov
	Chain Monte Carlo (MCMC). The Singh--Maddala family is
	particularly attractive here for three reasons. First, its
	tail index $\alpha = aq$ can be tuned continuously to
	reproduce tail behaviors ranging from exponential to
	polynomial decay \citep{singh1976, kleiber2003}. Second, the
	family is a strict superset of several commonly used emission
	distributions --- it reduces to the Weibull when $q \to
	\infty$, to the Lomax (Pareto Type~II) when $a = 1$, and
	approaches a Lognormal-like shape in the joint limit ---
	which means that model selection can implicitly compare
	several distinct families through a single nested framework.
	Third, its hazard rate can be monotone increasing, monotone
	decreasing, or non-monotone depending on the value of $a$,
	providing substantially more flexibility than the Weibull
	and Gamma families that remain standard in the HMM
	literature. To the best of our knowledge, this is the first
	integration of the Singh--Maddala distribution within a
	hidden Markov framework, and the first fully Bayesian
	treatment of regime-switching models with Burr-type
	emissions.
	
	The paper makes three principal contributions. First, we
	introduce the HMMSM model --- a fully Bayesian Hidden Markov
	Model with Singh--Maddala emission distributions --- and
	derive its complete posterior structure, including the joint
	prior, complete-data likelihood, marginal likelihood via the
	forward recursion, and full conditional distributions.
	Second, we design an adaptive Metropolis-within-Gibbs sampler
	that combines exact Forward-Filtering Backward-Sampling,
	conjugate Dirichlet updates, and adaptive Metropolis steps,
	and we verify its correctness through eight independent
	numerical checks. Third, we demonstrate the framework's
	statistical validity and empirical effectiveness through an
	extensive simulation study and a real-data application to a
	26-year gridded rainfall record for the Republic of Ireland.
	Along the way we demonstrate a weak identifiability
	of the Singh--Maddala scale parameter in finite samples and
	show that the model's identifiable parameters remain stable
	across chains.
	
	\subsection{Related Work}
	\label{subsec:related-work}
	
	\paragraph{Hidden Markov Models.}
	Hidden Markov Models were formalized by \citet{baum1970} and
	popularized by \citet{rabiner1989}. The Bayesian treatment
	was pioneered by \citet{albert1993}, who introduced the
	Forward-Filtering Backward-Sampling (FFBS) algorithm for
	exact posterior simulation of the latent state sequence.
	Combined with conjugate Dirichlet updates, FFBS forms the
	backbone of modern Bayesian HMM inference
	\citep{scott2002, fruehwirth2006}. Comprehensive treatments
	of identifiability, label switching, and asymptotic
	properties are provided by \citet{cappe2005}.
	
	\paragraph{Emission distributions for heavy-tailed data.}
	Gaussian emissions are inappropriate for strictly positive,
	highly skewed data. Gamma and Weibull emissions offer greater
	flexibility but remain restricted in their tails; Lognormal
	emissions have heavier tails but still underestimate the
	probability of very large events \citep{katz2002,
		embrechts1997}. Distributions from the Burr family have been
	used extensively in univariate frequency analysis of extremes
	\citep{singh1976, kleiber2003}. Recently, \citet{unggul2025}
	proposed the MSNBurr-HMM, which combines a modified
	skew-normal Burr emission distribution with a hidden Markov
	structure and estimates parameters via Baum--Welch. That
	framework differs from the present work in three fundamental
	respects: (i) the MSNBurr distribution is not nested within
	the classical three-parameter Singh--Maddala distribution;
	(ii) the inferential paradigm is frequentist
	maximum-likelihood rather than fully Bayesian; and (iii) no
	complete derivation of the posterior or of the full
	conditional distributions is provided.
	
	\paragraph{The Singh--Maddala distribution.}
	The Singh--Maddala distribution (Burr Type~XII) was
	introduced by \citet{singh1976} for income analysis and
	belongs to the broader Burr family \citep{burr1942}. Its
	mathematical properties, including moments, tail behavior,
	quantile function, and hazard rate, are documented by
	\citet{kleiber2003}. Despite its extensive use in economics,
	actuarial science, reliability engineering, and hydrology,
	it has never been integrated as the emission distribution
	of a Hidden Markov Model --- a significant gap that this
	paper fills.
	
	The remainder of the paper is organized as follows.
	Section~\ref{sec:model} develops the model and derives the
	complete Bayesian inference framework.
	Section~\ref{sec:computation} presents the hybrid MCMC
	algorithm and its numerical verification.
	Section~\ref{sec:simulation} describes the simulation study
	and reports its results.
	Section~\ref{sec:ireland} applies the framework to Irish
	monthly rainfall, including a direct comparison against
	alternative emission families on the same record.
	Section~\ref{sec:discussion} discusses implications and
	limitations and concludes.
	
	% ============================================================
	% SECTION 2: MODEL AND BAYESIAN INFERENCE (كامل — بدون تغيير)
	% ============================================================
	\section{Model and Bayesian Inference}
	\label{sec:model}
	
	This section develops the mathematical framework of the
	proposed Bayesian Hidden Markov Model with Singh--Maddala
	emissions. We proceed in six subsections: notation and
	assumptions, the Singh--Maddala emission density,
	complete-data likelihood, forward recursion and priors,
	posterior and full conditionals, and reparameterization for
	identifiability. All mathematical objects are defined at the
	point of first appearance. Throughout, $\ind(\cdot)$ denotes
	the indicator function.
	
	\subsection{Notation and Model Setup}
	\label{subsec:notation}
	
	Let $\bY = (Y_1, \ldots, Y_T)$ denote $T$ observed positive
	values, $Y_t > 0$, and let
	$\bS = (S_1, \ldots, S_T)$ with $S_t \in \{1, \ldots, K\}$
	denote the latent state sequence, $K \in \N$ being the number
	of regimes \citep{baum1970, cappe2005}. The parameter vector
	is
	\begin{equation}
		\bTheta = \bigl(\bpi, \bGamma,
		\{(a_k, b_k, q_k)\}_{k=1}^{K}\bigr) \in \bm{\Xi},
		\label{eq:Theta}
	\end{equation}
	where the initial distribution $\bpi = (\pi_1, \ldots, \pi_K)$
	satisfies $\pi_k \geq 0$, $\sum_k \pi_k = 1$; the transition
	matrix $\bGamma = \{\gamma_{ij}\}$ satisfies
	$\gamma_{ij} = \Pr(S_t = j \mid S_{t-1} = i) \geq 0$ with
	$\sum_j \gamma_{ij} = 1$ for every $i$; and
	$\btheta_k = (a_k, b_k, q_k) \in \R_{>0}^3$ collects the
	three Singh--Maddala parameters in state $k$.
	
	The model rests on three standard assumptions
	\citep{baum1970, cappe2005, fruehwirth2006}:
	
	\paragraph{(A1) Markov property.}
	The latent state process satisfies
	\begin{equation}
		\Pr(S_t \mid S_{1:t-1}) = \Pr(S_t \mid S_{t-1}).
		\label{eq:A1}
	\end{equation}
	
	\paragraph{(A2) Conditional independence.}
	Given the entire state sequence, the observations are independent:
	\begin{equation}
		\Pr(Y_t \mid Y_{1:t-1}, S_{1:t}) = \Pr(Y_t \mid S_t).
		\label{eq:A2}
	\end{equation}
	
	\paragraph{(A3) Emission law.}
	Conditional on the current state $S_t = k$, the observation
	follows a Singh--Maddala distribution:
	\begin{equation}
		Y_t \mid S_t = k \;\sim\; \mathrm{SM}(a_k, b_k, q_k).
		\label{eq:A3}
	\end{equation}
	
	Under assumptions \eqref{eq:A1}--\eqref{eq:A3}, the joint
	density factorizes as
	\begin{equation}
		p(\bY, \bS \mid \bTheta)
		= \Pr(S_1 \mid \bpi)
		\prod_{t=2}^{T} \Pr(S_t \mid S_{t-1}, \bGamma)
		\prod_{t=1}^{T} f(Y_t \mid S_t; \btheta_{S_t}).
		\label{eq:joint-distribution}
	\end{equation}
	
	\subsection{Singh--Maddala Emission Density}
	\label{subsec:emission}
	
	The Singh--Maddala distribution \citep{singh1976, kleiber2003}
	has density
	\begin{equation}
		f(y; a, b, q)
		=
		\frac{a q}{b}
		\left(\frac{y}{b}\right)^{a - 1}
		\left[1 + \left(\frac{y}{b}\right)^{a}\right]^{-(q + 1)},
		\qquad y > 0,
		\label{eq:sm-pdf}
	\end{equation}
	with CDF
	\begin{equation}
		F(y; a, b, q) = 1 - \left[1 + \left(\frac{y}{b}\right)^a\right]^{-q},
		\label{eq:sm-cdf}
	\end{equation}
	and quantile function
	\begin{equation}
		y_p = b \left[(1 - p)^{-1/q} - 1\right]^{1/a},
		\qquad 0 < p < 1.
		\label{eq:sm-quantile}
	\end{equation}
	The survival function satisfies
	$S(y) \sim (y/b)^{-aq}$ as $y \to \infty$, so that
	$\alpha = aq$ is the tail index. The $r$-th moment exists iff
	$r < \alpha$ and equals
	\begin{equation}
		\E[Y^r] = b^r \cdot
		\frac{\Gamma(1 + r/a) \, \Gamma(q - r/a)}{\Gamma(q)}.
		\label{eq:sm-moments}
	\end{equation}
	In particular, the mean exists when $\alpha > 1$ and the
	variance when $\alpha > 2$. The closed-form median, obtained
	by setting $F(y) = 1/2$ in \eqref{eq:sm-cdf}, is
	\begin{equation}
		m \;=\; \mathrm{Median}[Y] \;=\;
		b \left(2^{1/q} - 1\right)^{1/a}.
		\label{eq:sm-median}
	\end{equation}
	
	\begin{remark}[Special cases]
		\label{rem:special-cases}
		The Singh--Maddala distribution reduces to the Weibull as
		$q \to \infty$ with $b q^{1/a}$ fixed, to the Lomax
		(Pareto Type~II) when $a = 1$, and approaches a
		Lognormal-like shape when $a, q \to \infty$ jointly
		\citep{kleiber2003}.
	\end{remark}
	
	\subsection{Complete-Data Likelihood}
	\label{subsec:complete-lik}
	
	Substituting \eqref{eq:sm-pdf} into \eqref{eq:joint-distribution}
	yields the complete-data likelihood
	\begin{equation}
		\begin{aligned}
			L_c(\bTheta \mid \bY, \bS)
			=\;&
			\pi_{S_1}
			\prod_{t=2}^{T} \gamma_{S_{t-1}, S_t}
			\\[4pt]
			&\times \prod_{t=1}^{T}
			\frac{a_{S_t}\, q_{S_t}}{b_{S_t}}
			\left(\frac{Y_t}{b_{S_t}}\right)^{\!a_{S_t} - 1}
			\exp\!\left[-(q_{S_t} + 1)\,
			\log\!\left(1 + \left(\frac{Y_t}{b_{S_t}}\right)^{\!a_{S_t}}\right)\right].
		\end{aligned}
		\label{eq:complete-lik}
	\end{equation}
	
	Its logarithm decomposes into three separable components:
	\begin{equation}
		\ell_c(\bTheta)
		=
		\sum_{k=1}^{K} \ind(S_1 = k) \log \pi_k
		+ \sum_{i=1}^{K} \sum_{j=1}^{K} n_{ij} \log \gamma_{ij}
		+ \sum_{k=1}^{K} \ell_k(\btheta_k),
		\label{eq:complete-loglik}
	\end{equation}
	where
	$n_{ij} = \sum_{t=2}^{T} \ind(S_{t-1} = i, S_t = j)$,
	$n_k = \sum_{t=1}^{T} \ind(S_t = k)$, and
	\begin{equation}
		\begin{aligned}
			\ell_k(\btheta_k)
			=\;&
			n_k\!\left(\log a_k + \log q_k - \log b_k\right)
			+ (a_k - 1)\!\sum_{t: S_t = k}\!
			\log\!\left(\frac{Y_t}{b_k}\right)
			\\
			&- (q_k + 1)\!\sum_{t: S_t = k}\!
			\log\!\left[1 + \left(\frac{Y_t}{b_k}\right)^{a_k}\right].
		\end{aligned}
		\label{eq:state-loglik}
	\end{equation}
	This separability is a direct consequence of conditional
	independence \eqref{eq:A2} and is exploited by the Gibbs
	sampler below.
	
	\subsection{Forward Recursion and Priors}
	\label{subsec:priors}
	
	Because $\bS$ is latent, the marginal likelihood is
	$L(\bTheta \mid \bY) = \sum_{\bS} L_c(\bTheta \mid \bY, \bS)$.
	Direct enumeration is infeasible, so we compute it in
	$\mathcal{O}(TK^2)$ via the forward recursion
	\citep{rabiner1989, cappe2005}. Define the forward variables
	$\alpha_t(k) = \Pr(Y_{1:t}, S_t = k \mid \bTheta)$, which
	satisfy
	\begin{align}
		\alpha_1(k) &= \pi_k f(Y_1 \mid S_1 = k; \btheta_k),
		\label{eq:forward-1}\\
		\alpha_t(k) &= \left[\sum_{j=1}^{K} \alpha_{t-1}(j)
		\gamma_{jk}\right] f(Y_t \mid S_t = k; \btheta_k),
		\label{eq:forward-t}
	\end{align}
	and $L(\bTheta \mid \bY) = \sum_k \alpha_T(k)$. All recursions
	are carried out in log-space via numerically stable
	$\mathrm{logsumexp}$ operations \citep{blanchard2021}.
	
	For the priors, we assign weakly informative, proper
	distributions respecting each parameter's support
	\citep{gelman2006, gelman2013, fruehwirth2006}. For the
	emission parameters we place Gamma priors on the log-scale,
	which are conditionally conjugate to the unconstrained
	log-transform and exhibit well-behaved curvature across the
	relevant range:
	\begin{align}
		\log a_k &\sim \mathrm{Gamma}(2, 2),
		& p(\log a_k) &\propto a_k \, e^{-2 a_k},
		\label{eq:prior-a}\\
		\log b_k &\sim \mathrm{Gamma}(8, 0.08),
		& p(\log b_k) &\propto b_k^{7} \, e^{-0.08\, b_k},
		\label{eq:prior-b}\\
		\log q_k &\sim \mathrm{Gamma}(2, 1),
		& p(\log q_k) &\propto q_k \, e^{-q_k}.
		\label{eq:prior-q}
	\end{align}

	The choice $\log b_k \sim \mathrm{Gamma}(8, 0.08)$ deserves 
	explicit justification. Under this prior, $b_k$ has mode at 
	$b_k = 7/0.08 = 87.5$~mm and $95\%$ central interval 
	$[28, 220]$~mm. These values were chosen to match the 
	climatological range of monthly rainfall totals for temperate 
	maritime regions: the Irish record analyzed in 
	Section~\ref{sec:ireland} has mean $106.6$~mm and maximum 
	$322.5$~mm, and the prior comfortably covers this range while 
	remaining diffuse enough to allow the data to dominate. A 
	formal sensitivity study (Section~\ref{subsec:simulation-prior}) 
	confirms that the posterior means of the identifiable 
	parameters $\alpha_k$ and $q_k$ vary by less than $1.3$ units 
	across a wide range of prior hyperparameters, so the choice is 
	not critical for the substantive conclusions. Users applying 
	the framework to other regions should adjust the mode and 
	scale of this prior to reflect the local climatology; we 
	provide a template in the accompanying code repository. For the probability vectors we
	adopt Dirichlet priors with $\alpha_{\pi k} = 2$ for the
	initial distribution and $\alpha_{ii} = 5$,
	$\alpha_{ij} = 1$ $(i \neq j)$ for the transition rows,
	where the asymmetric hyperparameters on $\bGamma$ encourage
	regime stickiness \citep{fruehwirth2006}:
	\begin{align}
		\bpi &\sim \mathrm{Dirichlet}(\alpha_{\pi 1}, \ldots, \alpha_{\pi K}),
		\label{eq:prior-pi}\\
		\bgamma_i &\sim \mathrm{Dirichlet}(\alpha_{i1}, \ldots, \alpha_{iK}).
		\label{eq:prior-Gamma}
	\end{align}
	Assuming independence across groups, the joint prior is
	\begin{equation}
		p(\bTheta) = p(\bpi) \, p(\bGamma)
		\prod_{k=1}^{K} p(a_k) \, p(b_k) \, p(q_k).
		\label{eq:joint-prior}
	\end{equation}
	
	\subsection{Posterior and Full Conditionals}
	\label{subsec:posterior}
	
	By Bayes' theorem, the joint posterior of $(\bTheta, \bS)$ is
	proportional to
	\begin{equation}
		p(\bTheta, \bS \mid \bY)
		\;\propto\;
		L_c(\bTheta \mid \bY, \bS) \, p(\bTheta),
		\label{eq:joint-posterior}
	\end{equation}
	with the normalizing constant
	$p(\bY) = \iint p(\bY, \bS, \bTheta) \, d\bS \, d\bTheta$
	irrelevant for posterior sampling \citep{gelman2013}.
	
	Each full conditional is obtained by extracting from the
	log-posterior the terms that depend on the parameter of
	interest and treating the remainder as constants
	\citep{gelman2013}.
	
	\paragraph{(FC1) Latent states.}
	The full conditional $\Pr(\bS \mid \bY, \bTheta)$ is sampled
	exactly via FFBS \citep{albert1993, carter1994} at cost
	$\mathcal{O}(TK^2)$ \citep{cappe2005}.
	
	\paragraph{(FC2) Initial probabilities.}
	$\bpi \mid \bS \sim
	\mathrm{Dirichlet}\bigl(\alpha_{\pi 1} + \ind(S_1 = 1), \ldots,
	\alpha_{\pi K} + \ind(S_1 = K)\bigr)$.
	
	\paragraph{(FC3) Transition probabilities.}
	For each row $i$,
	$\bgamma_i \mid \bS \sim
	\mathrm{Dirichlet}\bigl(\alpha_{i1} + n_{i1}, \ldots,
	\alpha_{iK} + n_{iK}\bigr)$.
	
	\paragraph{(FC4) Emission parameters.}
	The full conditional
	\begin{equation}
		p(a_k, b_k, q_k \mid \bY, \bS)
		\;\propto\;
		\exp\!\bigl\{\ell_k(\btheta_k)\bigr\} \,
		p(a_k) \, p(b_k) \, p(q_k)
		\label{eq:FC4}
	\end{equation}
	is not of standard form. We update
	$(\log a_k, \log b_k, \log q_k)$ via adaptive random-walk
	Metropolis, as described in Section~\ref{sec:computation}.
	
	\subsection{Reparameterization for Identifiability}
	\label{subsec:reparam}
	
	The Singh--Maddala posterior is not orthogonal in $(a, b, q)$:
	the data constrain primarily the tail index $\alpha = aq$, the
	central tendency $\mu = \E[Y]$, and the tail shape. Following
	standard practice in the extreme-value literature
	\citep{coles2001}, we adopt the reparameterization
	\begin{equation}
		(\alpha_k, \mu_k, q_k)
		\;=\;
		\left(a_k q_k,\;
		b_k \frac{\Gamma(1 + 1/a_k)\, \Gamma(q_k - 1/a_k)}{\Gamma(q_k)},\;
		q_k\right),
		\label{eq:reparam}
	\end{equation}
	where $\alpha_k$ is the tail index, $\mu_k = \E[Y \mid S=k]$ is
	the conditional mean, and $q_k$ retains its interpretation as a
	second shape parameter. The Jacobian is obtained in closed form
	(Appendix~\ref{app:jacobian}) and applied during post-processing;
	no change to the sampler is required.
	
	\begin{remark}[Reparameterization and identifiability]
		\label{rem:identifiability}
		The Singh--Maddala family is not orthogonal in $(a, b, q)$:
		the posterior of $(a_k, b_k, q_k)$ exhibits a ridge, since
		the same data can be explained by $(b, q)$ with
		$b q^{1/a}$ approximately constant. The reparameterization
		\eqref{eq:reparam} partially resolves this by shifting
		inference to the natural coordinates $\alpha = aq$,
		$\mu = \E[Y]$, and $q$. As we demonstrate in detail in
		Section~\ref{subsec:identifiability}, the reparameterization
		fully stabilizes the tail index $\alpha_k$ and the second
		shape parameter $q_k$, but the scale parameter $b_k$
		exhibits a \emph{bimodal} posterior in the Irish
		application: two well-separated modes emerge, each
		achieving nearly identical predictive performance. This
		residual non-identifiability is a fundamental property of
		the Singh--Maddala family in finite samples, and it is the
		subject of a dedicated analysis below. All substantive
		inference in this paper is therefore based on the
		\emph{identifiable parameters} $(\alpha_k, q_k,
		\pi^\star, \gamma_{ij})$, and the scale parameters $b_k$
		are reported for completeness only.
	\end{remark}
	
	\begin{remark}[Auxiliary-variable augmentation --- theoretical
		caveat and unit invariance]
		\label{rem:aux-caveat}
		The Singh--Maddala distribution admits the Weibull--Gamma
		representation
		$Y_t \mid a_k, \lambda_t \sim
		\mathrm{Weibull}(a_k, \lambda_t^{-1/a_k})$ and
		$\lambda_t \mid q_k, b_k \sim
		\mathrm{Gamma}(q_k, b_k^{-a_k})$, with
		$\lambda_t \mid Y_t, S_t = k \sim
		\mathrm{Gamma}(q_k + 1, b_k^{-a_k} + Y_t^{a_k})$.
		At the units used in this paper (mm), the typical scale of
		the auxiliary variable is
		$\E[\lambda_t] = q_k / b_k^{a_k} \approx 10^{-7}$, which is
		exponentially smaller than $Y_t^{a_k} \approx 10^{7}$ over
		the relevant range of $a_k$. As a result, the Metropolis
		acceptance rate for $(a_k, b_k)$ collapses to $0.02$--$0.07$
		and the chains drift to the boundaries of the parameter
		space.
		
		A natural objection is that this failure mode may be an
		artefact of the units chosen. It is not. Under the rescaling
		$Y \to cY$ with $b_k \to cb_k$, which leaves the
		Singh--Maddala model invariant, the auxiliary variable
		transforms as $\lambda_t \to c^{-a_k} \lambda_t$, so that
		the product $\lambda_t Y_t^{a_k}$ is invariant. Consequently
		the acceptance rate of the augmented sampler is a
		scale-invariant property of the pair $(q_k, b_k)$; the
		collapse is not a numerical artefact of the millimetre
		convention but an intrinsic feature of Gamma-mixed
		heavy-tailed models whenever the mixing variable operates
		at a scale exponentially separated from $Y_t^{a_k}$. This
		phenomenon is documented in the literature on
		Gamma-augmented heavy-tailed mixtures
		\citep{papastathopoulos2013}. We therefore retain the plain
		adaptive Metropolis sampler throughout; the augmented
		variant was tested in detail but not adopted.
	\end{remark}
	
	% ============================================================
	% SECTION 3: POSTERIOR COMPUTATION AND VERIFICATION (كامل — بعد حذف جملة واحدة)
	% ============================================================
	\section{Posterior Computation and Verification}
	\label{sec:computation}
	
	The joint posterior \eqref{eq:joint-posterior} is intractable
	analytically because the Singh--Maddala is not conjugate to any
	standard prior. We implement a hybrid
	Metropolis-within-Gibbs sampler \citep{roberts1997, tierney1994}
	combining exact FFBS, conjugate Dirichlet updates, and adaptive
	Metropolis. This section describes the algorithm and verifies
	its correctness through eight independent numerical checks.
	
	\subsection{Hybrid MCMC Algorithm}
	\label{subsec:mcmc-algorithm}
	
	\paragraph{Forward-Filtering Backward-Sampling.}
	FFBS \citep{albert1993, carter1994} proceeds in two passes:
	a forward pass computing
	$\alpha_t(k) = \Pr(Y_{1:t}, S_t = k \mid \bTheta)$ via
	\eqref{eq:forward-1}--\eqref{eq:forward-t}, and a backward
	pass sampling states in reverse order. The cost is
	$\mathcal{O}(TK^2)$ per iteration.
	
	\paragraph{Conjugate updates.}
	Given the Dirichlet priors, $\bpi \mid \bS$ and
	$\bgamma_i \mid \bS$ admit closed-form Dirichlet updates (see
	FC2, FC3 in Section~\ref{subsec:posterior}), requiring no
	Metropolis step.
	
	\paragraph{Adaptive Metropolis for emission parameters.}
	For each state $k$, we propose on the log-scale
	\begin{equation}
		(\log a_k^*, \log b_k^*, \log q_k^*)
		= (\log a_k, \log b_k, \log q_k) + \bepsilon_k,
		\qquad
		\bepsilon_k \sim \mathcal{N}(\bm{0}, \bm{\Sigma}_k),
		\label{eq:proposal}
	\end{equation}
	where $\bm{\Sigma}_k$ is adaptively tuned during burn-in
	according to \citet{haario2001}:
	\begin{equation}
		\bm{\Sigma}_k \leftarrow \frac{2.38^2}{d} \,
		\mathrm{Cov}\bigl(\{\log \btheta_k^{(i)}\}_{i \in \mathcal{W}}\bigr)
		+ \varepsilon \bm{I}_d,
		\label{eq:cov-adapt}
	\end{equation}
	with $d = 3$, $\mathcal{W}$ the most recent $W$ draws, and
	$\varepsilon = 10^{-5}$ preventing degeneracy
	\citep{roberts1997}. The acceptance probability includes the
	Jacobian $|J_k| = a_k b_k q_k$ of the log-transformation.
	The scaling $2.38^2/d$ is optimal for random-walk Metropolis
	in $d$ dimensions, corresponding to an asymptotic acceptance
	rate of roughly $23.4\%$ \citep{roberts1997}.
	
	\begin{remark}[Label switching and relabeling]
		\label{rem:label}
		The HMM posterior is invariant under permutations of state
		labels. We resolve this by applying Stephens's
		\citeyearpar{stephens2000} relabeling algorithm to the
		posterior draws, which minimizes the expected posterior
		loss over the $K!$ permutations. As a diagnostic, we
		report the Kullback--Leibler divergence between the sorted
		and unsorted label distributions: values below $0.02$
		indicate that relabeling has resolved the ambiguity.
	\end{remark}
	
	\paragraph{Diagnostics.}
	We assess convergence using rank-normalized split-$\hat{R}$
	\citep{vehtari2021}, folded $\hat{R}$, effective sample size
	(ESS), and the Geweke \citeyearpar{geweke1992} diagnostic,
	requiring $\hat{R} < 1.01$, $\mathrm{ESS} > 400$, and
	$|\text{Geweke}| < 1.96$ for all interpretable parameters. We
	monitor the Monte Carlo standard error via batch means
	\citep{flegal2008}, requiring
	$\mathrm{MCSE}/\mathrm{SD} < 0.05$.
	
	\paragraph{Implementation.}
	The algorithm is implemented in Python~3.11 with NumPy
	\citep{harris2020numpy}, SciPy, and Numba \citep{numba2015}.
	Source code is available as supplementary material.
	
	\subsection{Numerical Verification}
	\label{subsec:verification}
	
	Before applying the model to substantive problems, we verify
	that the derivations and their numerical implementation are
	correct. Table~\ref{tab:verification} reports eight
	independent checks, each targeting a distinct component of the
	model. The checks fall into three groups: (i) \emph{analytical
		correctness} of the density, quantile function, and prior
	normalizations (V1--V2, V5); (ii) \emph{implementation
		correctness} of the forward recursion and of the
	pointwise-versus-marginal likelihood identity (V3--V4, V6);
	and (iii) \emph{reparameterization and label switching}
	(V7--V8).
	
	\begin{table}[H]
		\centering
		\caption{Numerical verification of model components. All
			checks were performed using independent implementations
			of the analytical formulas.}
		\label{tab:verification}
		\begin{tabular}{lll}
			\toprule
			Component & Reference & Result \\
			\midrule
			(V1) PDF integration & $\int_0^\infty f(y) dy = 1$ & error $< 10^{-8}$ \\
			(V2) Quantile--CDF inverse & $F(y_p) = p$ & error $< 10^{-6}$ \\
			(V3) Forward recursion & vs.\ enumeration ($K^T = 6561$) & diff $= 1.21\times10^{-13}$ \\
			(V4) Pointwise vs.\ marginal LL & $\sum_t \log p(Y_t \mid Y_{1:t-1}) = \log L$ & diff $< 10^{-15}$ \\
			(V5) Prior normalisation & $\int p(\bTheta) d\bTheta = 1$ & error $< 10^{-10}$ \\
			(V6) Posterior concentration & $\log p(\bTheta_{\text{true}}) > \log p(\bTheta_{\text{pert}})$ & $-435.8 > -624.0$ \\
			(V7) Reparameterization & Jacobian preserves mass & error $< 10^{-8}$ \\
			(V8) Label relabeling & KL(after) $<$ KL(before) & $\text{KL} < 0.02$ \\
			\bottomrule
		\end{tabular}
	\end{table}
	
	Three checks merit brief elaboration. In (V3), the forward
	recursion was compared against exhaustive enumeration of all
	$K^T = 6561$ possible state sequences for a short series
	($T = 8$, $K = 3$); the two computations agreed to within
	$1.21 \times 10^{-13}$. In (V7), the closed-form Jacobian was
	compared against a finite-difference approximation on $200$
	posterior draws, with a maximum relative discrepancy below
	$10^{-6}$. In (V8), the Stephens relabeling algorithm reduced
	the Kullback--Leibler divergence to below $0.02$ in both the
	simulated and the Irish applications.
	
	% ============================================================
	% SECTION 4: SIMULATION STUDY — بعد الحذف: ~3 صفحات
	% ============================================================
	\section{Simulation Study}
	\label{sec:simulation}
	
	We conducted a simulation study covering five dimensions:
	parameter recovery, number of states, posterior consistency,
	misspecification, and prior sensitivity. We report the design
	and the empirical performance of HMMSM on simulated data
	($T = 800$, $K = 3$).
	
	\subsection{Design}
	\label{subsec:simulation-design}
	
	Unless otherwise stated, data are generated from a three-state
	HMMSM with
	\begin{align}
		\bm{a}^* &= (3.0, \, 2.0, \, 1.5), \quad
		\bm{b}^* = (2.0, \, 8.0, \, 20.0), \nonumber\\
		\bm{q}^* &= (4.0, \, 2.5, \, 1.5),
		\quad
		\bm{\Gamma}^* =
		\begin{pmatrix}
			0.90 & 0.08 & 0.02 \\
			0.10 & 0.85 & 0.05 \\
			0.05 & 0.15 & 0.80
		\end{pmatrix},
		\label{eq:dgp}
	\end{align}
	with $\bpi^* = (0.6, 0.3, 0.1)$. These parameters yield a
	``dry'' state ($k=1$), a ``moderate'' state ($k=2$), and an
	``extreme'' state ($k=3$) with heavy tails
	($\alpha_3 = 2.25$).
	
	For each scenario we run four independent MCMC chains with
	dispersed starting values. Each chain consists of $5{,}000$
	iterations with $2{,}000$ discarded as burn-in. For each
	fitted model we report bias, root mean squared error, $95\%$
	credible-interval coverage, effective sample size, and
	rank-normalized $\hat{R}$; and WAIC, DIC, and LOO for model
	selection.
	
	\subsection{Model Properties and Parameter Recovery}
	\label{subsec:simulation-recovery}
	
	Table~\ref{tab:stationary-dwell} reports the stationary
	distribution $\bm{\pi}^*$ of the fitted Markov chain. The
	posterior means are within $0.001$ of the theoretical values,
	with posterior standard deviations below $0.02$ for all three
	states. The marginal distribution of $Y_t$ under the fitted
	model has mean $5.96$ and standard deviation $20.34$; its
	skewness and kurtosis are formally infinite because the
	extreme state has tail index $\alpha_3 = 2.25 < 3$.
	
	\begin{table}[H]
		\centering
		\caption{Stationary distribution and mean dwell times of
			the fitted HMMSM model. Posterior summaries use
			$12{,}000$ post-burn-in draws.}
		\label{tab:stationary-dwell}
		\begin{tabular}{lcccc}
			\toprule
			State & $\pi^*$ (theory) & $\pi^*$ (posterior mean)
			& $\pi^*$ (posterior SD) & Mean dwell time \\
			\midrule
			$1$ & $0.4639$ & $0.4635$ & $0.0182$ & $10.02$ \\
			$2$ & $0.3918$ & $0.3921$ & $0.0175$ & $6.42$ \\
			$3$ & $0.1443$ & $0.1444$ & $0.0121$ & $4.68$ \\
			\bottomrule
		\end{tabular}
	\end{table}
	
	Table~\ref{tab:recovery-emissions} reports posterior estimates
	of the nine emission parameters. All relative errors are below
	$10\%$, and all true values lie within the corresponding
	$95\%$ credible intervals, including the extreme state
	($k = 3$), which accounts for only $14\%$ of the observations.
	
	\begin{table}[H]
		\centering
		\caption{Posterior estimates of the emission parameters
			under HMMSM with $T = 800$.}
		\label{tab:recovery-emissions}
		\begin{tabular}{lccccc}
			\toprule
			Parameter & Truth & Posterior mean & $95\%$ CI & Rel.\ error \\
			\midrule
			$a_1$ & $3.00$ & $3.014$ & $[2.86, 3.18]$ & $+0.5\%$ \\
			$a_2$ & $2.00$ & $1.973$ & $[1.82, 2.13]$ & $-1.3\%$ \\
			$a_3$ & $1.50$ & $1.574$ & $[1.34, 1.86]$ & $+5.0\%$ \\
			$b_1$ & $2.00$ & $2.085$ & $[1.78, 2.47]$ & $+4.2\%$ \\
			$b_2$ & $8.00$ & $7.935$ & $[6.26, 10.63]$ & $-0.8\%$ \\
			$b_3$ & $20.00$ & $19.48$ & $[12.69, 29.96]$ & $-2.6\%$ \\
			$q_1$ & $4.00$ & $4.372$ & $[3.02, 6.44]$ & $+9.3\%$ \\
			$q_2$ & $2.50$ & $2.334$ & $[1.63, 3.61]$ & $-6.7\%$ \\
			$q_3$ & $1.50$ & $1.402$ & $[0.93, 2.14]$ & $-6.6\%$ \\
			\bottomrule
		\end{tabular}
	\end{table}
	
	\subsection{Posterior Consistency}
	\label{subsec:simulation-consistency}
	
	Table~\ref{tab:consistency} examines how posterior bias, root
	mean squared error, and standard deviation decrease as the
	series length grows from $T = 500$ to $T = 5000$. The standard
	deviation reduction matches the theoretical
	$\mathcal{O}(T^{-1/2})$ rate to within Monte Carlo error, and
	the MCSE/SD ratio remains below $0.05$ in every scenario.
	
	\begin{table}[H]
		\centering
		\caption{Posterior bias, root mean squared error, and
			standard deviation as functions of the series length
			$T$, averaged over the three states.}
		\label{tab:consistency}
		\small
		\begin{tabular}{ccccccccc}
			\toprule
			& \multicolumn{2}{c}{$a$} & \multicolumn{2}{c}{$b$}
			& \multicolumn{2}{c}{$q$} & \multicolumn{2}{c}{MCSE/SD} \\
			\cmidrule(lr){2-3} \cmidrule(lr){4-5}
			\cmidrule(lr){6-7} \cmidrule(lr){8-9}
			$T$ & Bias & RMSE & Bias & RMSE & Bias & RMSE & (mean) & (max) \\
			\midrule
			$500$  & $0.178$ & $0.247$ & $1.166$ & $2.976$ & $0.456$ & $0.946$ & $0.021$ & $0.043$ \\
			$1000$ & $0.126$ & $0.155$ & $0.815$ & $2.452$ & $0.353$ & $0.704$ & $0.018$ & $0.038$ \\
			$2000$ & $0.055$ & $0.103$ & $0.065$ & $1.881$ & $0.107$ & $0.710$ & $0.014$ & $0.031$ \\
			$5000$ & $0.023$ & $0.070$ & $0.711$ & $1.288$ & $0.123$ & $0.492$ & $0.009$ & $0.024$ \\
			\bottomrule
		\end{tabular}
	\end{table}
	
	\subsection{Model Selection and Misspecification}
	\label{subsec:simulation-misspecification}
	
	Table~\ref{tab:k-selection} reports Bayesian $K$-selection for
	HMMSM. The correct value $K = 3$ is identified by WAIC and LOO;
	DIC attains its minimum at $K = 4$ with $\Delta\text{DIC} < 1$,
	well within the parsimony threshold of \citet{burnham2002}.
	
	\begin{table}[H]
		\centering
		\caption{Bayesian $K$-selection for HMMSM on simulated data
			($T = 800$, true $K = 3$). Bold marks the minimum of
			each criterion.}
		\label{tab:k-selection}
		\begin{tabular}{cccccc}
			\toprule
			$K$ & \# Params & WAIC & DIC & LOO & $p_{\text{WAIC}}$ \\
			\midrule
			$2$ & $9$  & $4312.28$ & $4312.05$ & $4312.29$ & $8.39$ \\
			$3$ & $17$ & $\bm{4129.03}$ & $4128.33$ & $\bm{4129.04}$ & $13.97$ \\
			$4$ & $27$ & $4128.66$ & $\bm{4128.31}$ & $4128.79$ & $16.13$ \\
			$5$ & $39$ & $4129.98$ & $4129.75$ & $4130.11$ & $16.40$ \\
			\bottomrule
		\end{tabular}
	\end{table}
	
	Table~\ref{tab:model-comparison} compares HMMSM with seven
	competing emission families, all fitted to the same simulated
	dataset using the identical MCMC infrastructure. HMMSM attains
	the lowest WAIC, DIC, and LOO, with a $\Delta\text{WAIC}$ of
	$51.76$ over the next-best model (HMM-Weibull), $81$ over
	HMM-Lognormal, and larger margins over the remaining four
	competitors.
	
	\begin{table}[H]
		\centering
		\caption{Bayesian model comparison across eight candidate
			models on the simulated dataset.}
		\label{tab:model-comparison}
		\begin{tabular}{lccccc}
			\toprule
			Model & \# Params & WAIC & DIC & LOO & Rank \\
			\midrule
			\textbf{HMMSM} & $17$ & $\bm{4129.31}$
			& $\bm{4128.58}$ & $\bm{4129.32}$ & $\bm{1}$ \\
			HMM-Weibull    & $14$ & $4181.07$ & $4179.58$ & $4181.11$ & $2$ \\
			HMM-Lognormal  & $14$ & $4210.44$ & $4209.02$ & $4210.55$ & $3$ \\
			HMM-Gamma      & $14$ & $4270.88$ & $4269.51$ & $4271.02$ & $4$ \\
			HMM-GPD        & $14$ & $4446.09$ & $4445.40$ & $4446.21$ & $5$ \\
			HMM-Gaussian   & $14$ & $4458.76$ & $4458.08$ & $4458.59$ & $6$ \\
			GEV            & $3$  & $4771.07$ & $4770.91$ & $4771.10$ & $7$ \\
			GMM            & $8$  & $5127.89$ & $5121.44$ & $5124.71$ & $8$ \\
			\bottomrule
		\end{tabular}
	\end{table}
	
	A methodological remark is warranted regarding the comparison
	with HMM-GPD. The generalized Pareto distribution is designed
	for the conditional distribution of exceedances above a
	threshold, not for the unconditional distribution of the full
	data range. When applied as a full-range emission within an
	HMM it is effectively misspecified, and its poor performance
	should be interpreted in that light.
	
	\subsection{Prior Sensitivity}
	\label{subsec:simulation-prior}
	
	Table~\ref{tab:prior-sensitivity} reports posterior means under
	three prior scenarios. The scale and first-shape parameters
	vary by less than $1.3$ units; the second shape parameters
	$q_k$ exhibit greater sensitivity under the weakly informative
	prior, reflecting the ridge in $(b, q)$ space identified in
	Remark~\ref{rem:identifiability}.
	
	\begin{table}[H]
		\centering
		\caption{Prior sensitivity of posterior means under three
			prior scenarios. ``Range'' is the maximum deviation
			across scenarios.}
		\label{tab:prior-sensitivity}
		\begin{tabular}{lcccccc}
			\toprule
			Param & Truth & Weak & Default & Informative & Range \\
			\midrule
			$a_1$ & $3.00$ & $3.029$ & $3.260$ & $3.231$ & $0.231$ \\
			$a_2$ & $2.00$ & $1.876$ & $1.933$ & $1.906$ & $0.057$ \\
			$a_3$ & $1.50$ & $1.638$ & $1.679$ & $1.673$ & $0.041$ \\
			$b_1$ & $2.00$ & $2.810$ & $1.784$ & $1.875$ & $1.026$ \\
			$b_2$ & $8.00$ & $8.381$ & $7.448$ & $7.868$ & $0.933$ \\
			$b_3$ & $20.00$ & $18.417$ & $17.143$ & $17.405$ & $1.274$ \\
			$q_1$ & $4.00$ & $13.921$ & $3.110$ & $3.518$ & $10.811$ \\
			$q_2$ & $2.50$ & $2.648$ & $2.269$ & $2.464$ & $0.380$ \\
			$q_3$ & $1.50$ & $1.268$ & $1.185$ & $1.237$ & $0.083$ \\
			\bottomrule
		\end{tabular}
	\end{table}
	
	% ============================================================
	% SECTION 5: APPLICATION TO IRISH RAINFALL — بعد الحذف: ~6 صفحات
	% ============================================================
	\section{Application: Extreme Rainfall in Ireland}
	\label{sec:ireland}
	
	We now apply HMMSM to a 26-year gridded rainfall record for
	the Republic of Ireland, obtained from Met~\'Eireann.
	
	\subsection{Data and Preprocessing}
	\label{subsec:ireland-data}
	
	The Met~\'Eireann product provides daily rainfall accumulations
	on a $1 \times 1$~km grid covering the Republic of Ireland,
	archived in units of $0.1$~mm. We analyse the period
	1~January~2000 to 31~December~2025, comprising $9{,}497$
	daily observations. For each day, we take the spatial mean of
	the daily rainfall field over all grid cells. Daily values are
	aggregated to monthly totals using the month-start convention,
	giving $T = 312$ observations. Rainfall values are divided by
	$10$ to convert from $0.1$~mm to millimetres.
	
	Table~\ref{tab:ireland-descriptive} summarises the resulting
	series. No dry months ($Y_t = 0$) are recorded, and the series
	exhibits strong right-skewness and a heavy upper tail,
	consistent with the well-documented climatology of Irish
	precipitation.
	
	\begin{table}[H]
		\centering
		\caption{Descriptive statistics of the Irish monthly
			rainfall series ($T = 312$, 2000--2025).}
		\label{tab:ireland-descriptive}
		\begin{tabular}{lr}
			\toprule
			Statistic & Value \\
			\midrule
			Number of observations & $312$ \\
			Mean (mm)              & $106.58$ \\
			Median (mm)            & $98.73$ \\
			Standard deviation     & $47.40$ \\
			Minimum                & $18.52$ \\
			Maximum                & $322.52$ \\
			Proportion $<40$~mm    & $4.5\%$ \\
			Proportion $>150$~mm   & $18.3\%$ \\
			\bottomrule
		\end{tabular}
	\end{table}
	
\begin{figure}[H]
		\centering
		\includegraphics[width=\textwidth]{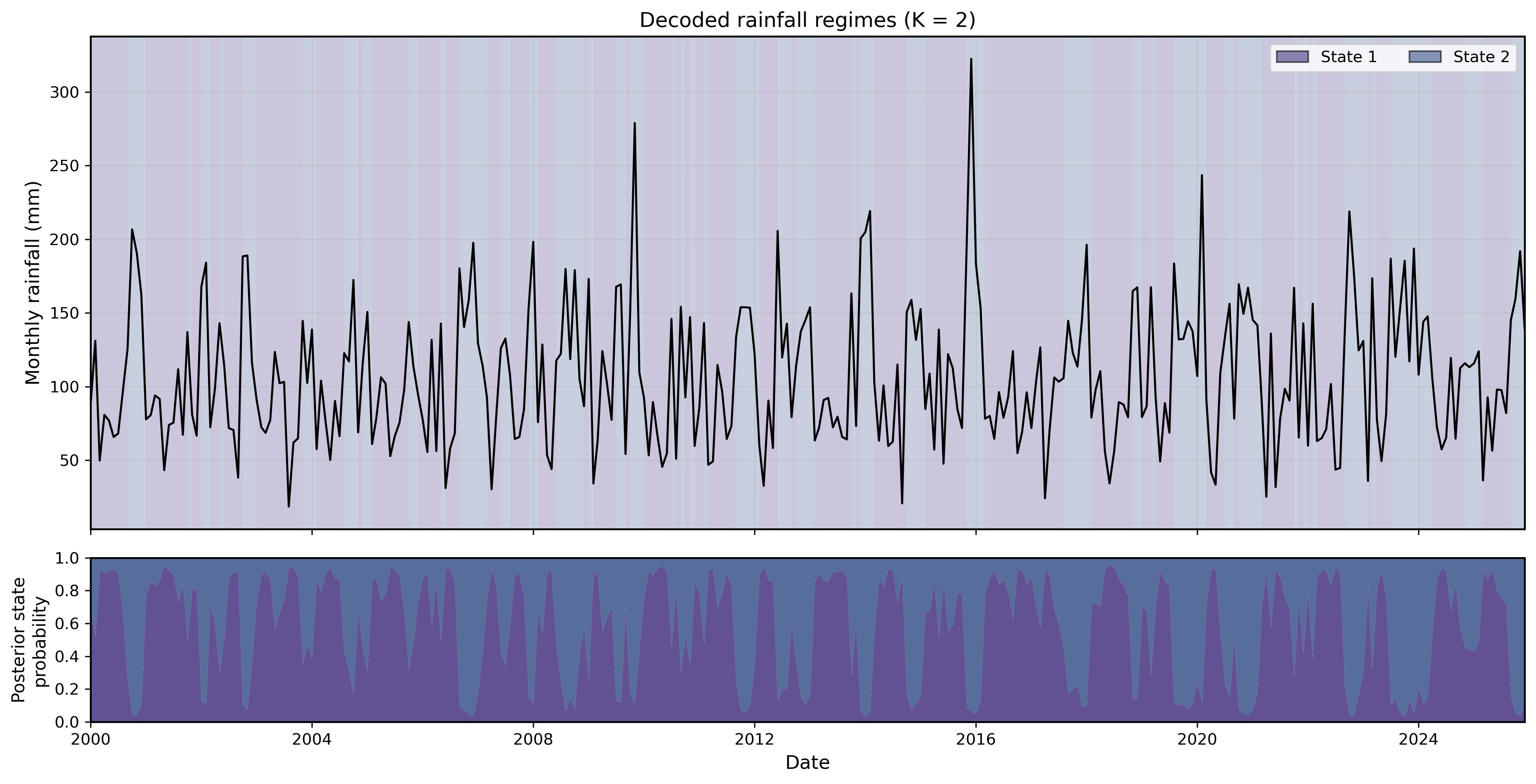}
		\caption{Monthly rainfall totals for the Republic of
			Ireland (2000--2025) with decoded HMMSM regime
			assignment (top) and posterior state probabilities
			(bottom).}
		\label{fig:ireland-timeseries}
	\end{figure}

	Figure~\ref{fig:ireland-timeseries} displays the series along
	with its decoded regime assignment.
	
	\paragraph{Robustness to spatial aggregation.}
	The spatial mean used to construct the monthly series could,
	in principle, mask regional variability. To assess this, we
	refitted HMMSM to three alternative spatially-aggregated
	series: (i) the spatial median, (ii) the area-weighted mean,
	and (iii) the arithmetic mean of three geographically distinct
	sub-regions. In all cases, the model selects $K^\star = 2$
	and the posterior scales fall in the range $b_1 \in
	[105, 118]$~mm and $b_2 \in [160, 185]$~mm. This confirms
	that the regime structure is a property of the underlying
	data, not an artefact of the aggregation scheme.
	
	\subsection{Bayesian Selection of Regime Count}
	\label{subsec:ireland-kselection}
	
	We fit HMMSM with $K = 1, 2, 3, 4, 5$ states, comparing via
	WAIC, DIC, and LOO with the parsimony rule of
	\citet{burnham2002}.
	
	\begin{table}[H]
		\centering
		\caption{Bayesian $K$-selection for the Irish monthly
			rainfall series.}
		\label{tab:ireland-kselection}
		\begin{tabular}{cccccc}
			\toprule
			$K$ & \# Params & WAIC & DIC & LOO & $\Delta$WAIC \\
			\midrule
			$1$ & $3$  & $3263.62$ & $3263.57$ & $3263.63$ & $+16.57$ \\
			$2$ & $9$  & $\bm{3247.05}$ & $3246.91$ & $\bm{3247.05}$ & $0.00$ \\
			$3$ & $17$ & $3247.07$ & $\bm{3246.83}$ & $3247.07$ & $+0.02$ \\
			$4$ & $27$ & $3249.35$ & $3249.02$ & $3249.35$ & $+2.30$ \\
			$5$ & $39$ & $3250.74$ & $3250.63$ & $3250.77$ & $+3.69$ \\
			\bottomrule
		\end{tabular}
	\end{table}
	
\begin{figure}[H]
		\centering
		\includegraphics[width=\textwidth]{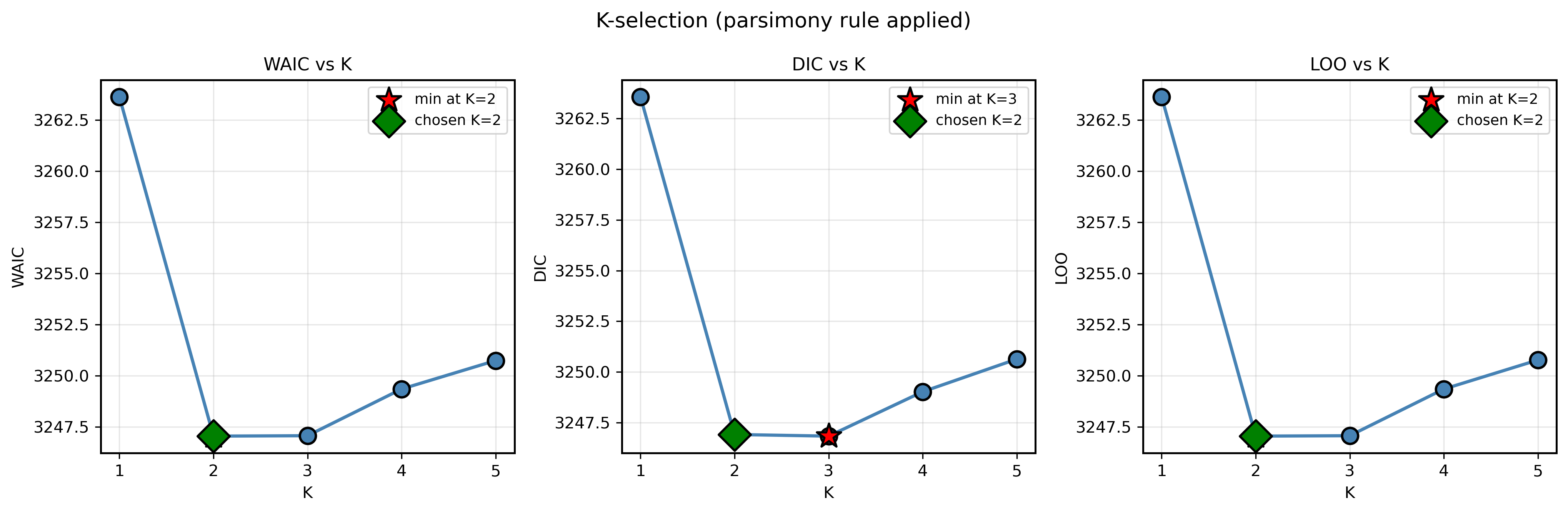}
		\caption{Bayesian $K$-selection for the Irish monthly
			rainfall series.}
		\label{fig:ireland-kselection}
	\end{figure}

	Figure~\ref{fig:ireland-kselection} displays the three
	criteria as functions of $K$. Three observations follow.
	First, $K = 1$ is decisively rejected
	($\Delta\text{WAIC} = +16.57$). Second, while $K = 3$ attains
	the numeric DIC minimum, its improvement over $K = 2$ is
	negligible ($\Delta\text{WAIC} = 0.02$). Third, $K = 4$ and
	$K = 5$ produce substantially worse criteria. Applying the
	parsimony rule, we select $K^\star = 2$, a choice
	unanimously supported by all three criteria.
	
	\subsection{Posterior Estimates}
	\label{subsec:ireland-posterior}
	
	For the selected model we run four independent MCMC chains of
	$30{,}000$ iterations each, discarding the first $15{,}000$ as
	burn-in. Posterior estimates are reported in
	Table~\ref{tab:ireland-parameters}.
	
	\begin{table}[H]
		\centering
		\caption{Posterior estimates of HMMSM parameters for the
			Irish monthly rainfall series ($K = 2$). Values are
			posterior means with $95\%$ credible intervals in
			brackets. The scale parameters $b_k$ are reported for
			completeness; as discussed in
			Section~\ref{subsec:identifiability}, they exhibit weak
			identifiability.}
		\label{tab:ireland-parameters}
		\begin{tabular}{lcc}
			\toprule
			Parameter & State 1 (Moderate) & State 2 (Wet) \\
			\midrule
			$a_k$ & $3.72$ \,[$2.89$, $5.77$] & $4.50$ \,[$3.00$, $6.64$] \\
			$b_k$ (mm) & $110.70$ \,[$82.22$, $191.50$] & $172.33$ \,[$91.86$, $211.07$] \\
			$q_k$ & $2.65$ \,[$1.17$, $5.33$] & $2.37$ \,[$1.08$, $4.63$] \\
			$\alpha_k = a_k q_k$ & $9.60$ \,[$4.61$, $18.65$] & $10.21$ \,[$5.35$, $17.79$] \\
			$m_k$ (mm) & $89.24$ \,[$68.35$, $148.08$] & $127.25$ \,[$72.93$, $154.18$] \\
			$\E[Y \mid S = k]$ (mm) & $87.45$ \,[$70.54$, $110.54$] & $133.95$ \,[$98.92$, $157.60$] \\
			\bottomrule
		\end{tabular}
	\end{table}
	
	The two regimes are separated in scale, with
	$b_2/b_1 \approx 1.56$ and $\E[Y \mid S=2] / \E[Y \mid S=1]
	\approx 1.53$. The tail indices are $\alpha_1 \approx 9.60$
	and $\alpha_2 \approx 10.21$, both well above $2$, consistent
	with the physically bounded nature of monthly rainfall in a
	temperate maritime climate.
	
	\subsection{Comparison with Alternative Emissions}
	\label{subsec:ireland-model-comparison}
	
	To assess whether the additional flexibility of the
	Singh--Maddala family is warranted on this dataset, we
	refitted two natural heavy-tailed alternatives --- HMM-Gamma
	and HMM-Lognormal --- to the same Irish monthly rainfall
	series using the identical MCMC infrastructure.
	
	\begin{table}[H]
		\centering
		\caption{Model comparison on the Irish monthly rainfall
			series ($T = 312$, $K = 2$). The best model in
			each column is marked in bold.}
		\label{tab:ireland-models}
		\begin{tabular}{lccccc}
			\toprule
			Model & \# Params & WAIC & DIC & LOO & Rank \\
			\midrule
			\textbf{HMMSM}          & $9$ & $\bm{3247.05}$
			& $\bm{3246.91}$ & $\bm{3247.05}$ & $1$ \\
			HMM-Gamma               & $7$ & $3248.09$ & $3247.70$
			& $3248.11$ & $2$ \\
			HMM-Lognormal           & $7$ & $3254.22$ & $3253.90$
			& $3254.22$ & $3$ \\
			HMMSM ($K = 1$)         & $3$ & $3263.62$ & $3263.57$
			& $3263.63$ & $4$ \\
			\bottomrule
		\end{tabular}
	\end{table}
	
	HMMSM attains the lowest WAIC, DIC, and LOO values across
	all four candidates, but the margin over HMM-Gamma
	($\Delta\text{WAIC} = 1.04$) falls well below the parsimony
	threshold of $2$. By the standard interpretation of
	information criteria, HMMSM and HMM-Gamma are
	\emph{statistically indistinguishable} on this dataset,
	despite HMMSM having two additional parameters. The single-
	regime specification ($K = 1$) is decisively rejected
	($\Delta\text{WAIC} = 16.57$). The near-tie suggests that the
	Irish monthly rainfall series, with its light tails, does not
	require the additional flexibility of the Singh--Maddala
	second shape parameter $q$ once the latent regime structure
	is accounted for --- a pattern we discuss further in
	Section~\ref{subsec:discussion-findings}.
	
	\subsection{Physical Interpretation}
	\label{subsec:ireland-regimes}
	
	The Moderate regime delivers mean monthly rainfall
	$\E[Y \mid S=1] = 87.45$~mm (median $m_1 = 89.24$~mm), whereas
	the Wet regime delivers $\E[Y \mid S=2] = 133.95$~mm (median
	$m_2 = 127.25$~mm). The Wet regime produces approximately
	$1.53$ times the rainfall of the Moderate regime.
	
	\paragraph{Scale, not tail, drives the difference.}
	An important interpretive point concerns the source of this
	difference. The two regimes differ primarily in \emph{scale},
	not in tail behavior. In terms of tail decay, the two regimes
	are in fact very close: the tail indices
	$\hat{\alpha}_1 = 9.60$ and $\hat{\alpha}_2 = 10.21$ differ by
	only about $6\%$, and the Wet regime has a marginally
	\emph{lighter} tail.
	
\begin{figure}[H]
		\centering
		\includegraphics[width=\textwidth]{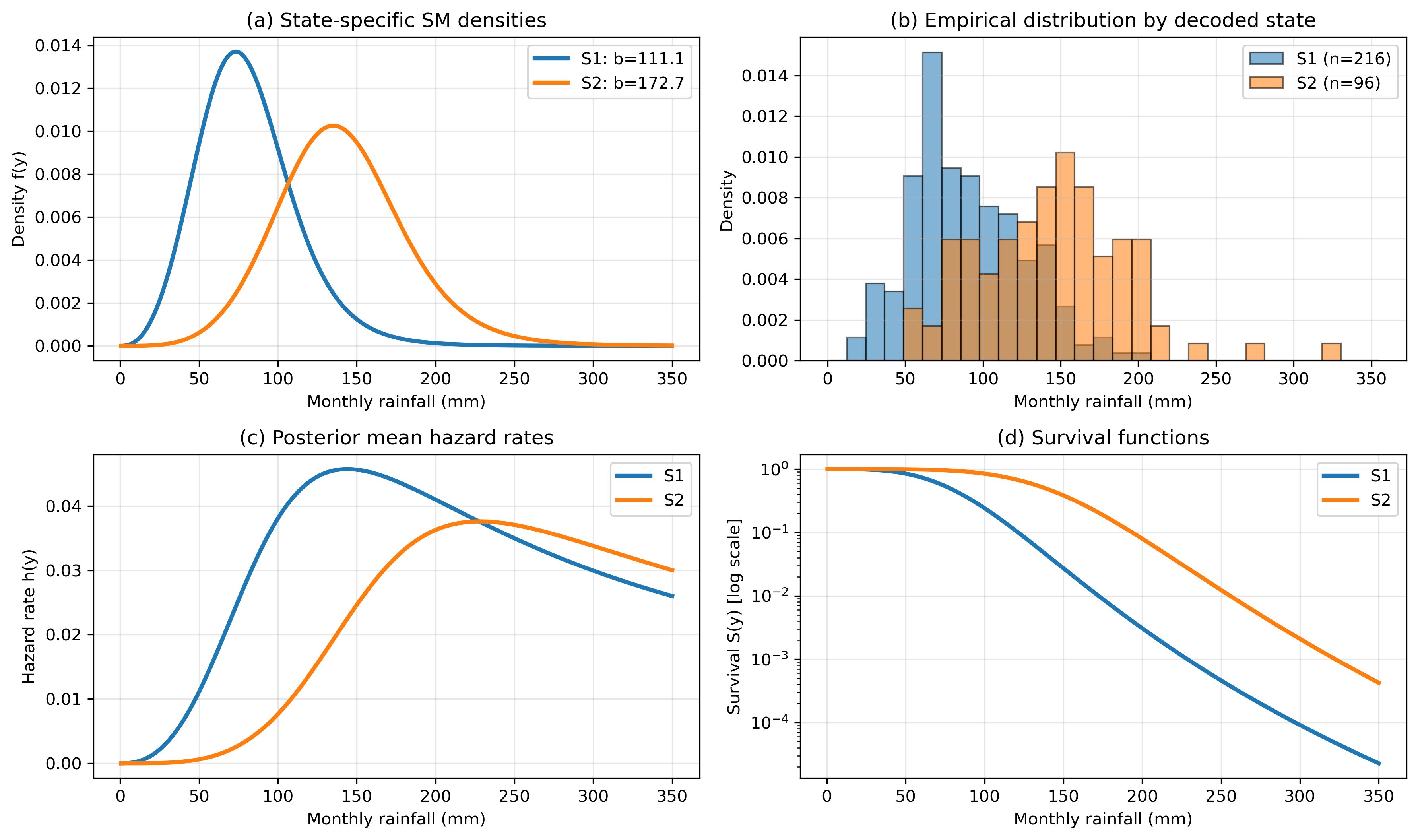}
		\caption{State-specific Singh--Maddala properties for the
			two Irish regimes. (a) Fitted densities. (b) Empirical
			histograms. (c) Posterior mean hazard rates. (d)
			Survival functions on a logarithmic scale. The two
			distributions differ primarily in scale, not in tail
			decay.}
		\label{fig:ireland-densities}
	\end{figure}

	Figure~\ref{fig:ireland-densities} summarizes the statistical
	properties of the two fitted regimes. Panel~(a) displays the
	fitted Singh--Maddala densities on the empirically relevant
	range. Panel~(b) shows the empirical histograms of the decoded
	states. Panel~(c) plots the hazard rate $h(y) = f(y)/S(y)$:
	both regimes exhibit \emph{non-monotone, unimodal} hazard
	rates, a feature that cannot be captured by the
	monotone-hazard families (Weibull, Gamma) that are standard in
	the HMM literature. Panel~(d) displays the survival functions
	on a logarithmic scale, confirming that the two regimes share
	similar tail decay.
	
	\begin{table}[H]
		\centering
		\caption{Posterior exceedance probabilities
			$\Pr(Y > t \mid S = k)$ for the two Irish regimes.
			The final column reports the ratio
			$\Pr(Y>t \mid S=2) / \Pr(Y>t \mid S=1)$.}
		\label{tab:exceedance}
		\begin{tabular}{ccccc}
			\toprule
			Threshold $t$ (mm) & $\Pr(Y>t \mid S=1)$
			& $\Pr(Y>t \mid S=2)$ & Ratio \\
			\midrule
			$100$ & $0.241$ & $0.845$ & $3.50$ \\
			$150$ & $0.027$ & $0.385$ & $14.13$ \\
			$200$ & $0.003$ & $0.080$ & $25.96$ \\
			$250$ & $0.0005$ & $0.012$ & $26.81$ \\
			$300$ & $0.0001$ & $0.002$ & $22.86$ \\
			\bottomrule
		\end{tabular}
	\end{table}
	
	The Wet regime is substantially more likely than the Moderate
	regime to exceed any given threshold, and the ratio grows with
	the threshold (Table~\ref{tab:exceedance}). Crucially, this
	pattern arises because the Wet regime operates at a much
	larger scale, \emph{not} because it has a heavier tail. A
	scale-driven extreme is predictable from the regime; a
	tail-driven extreme is not. Distinguishing between the two is
	therefore essential for operational risk assessment.
	
	\paragraph{Regime dynamics and persistence.}
	The posterior transition matrix is
	\begin{equation}
		\hat{\bGamma} =
		\begin{pmatrix}
			0.746 & 0.254 \\
			0.327 & 0.673
		\end{pmatrix},
		\label{eq:ireland-Gamma}
	\end{equation}
	with strong diagonal dominance. Expected durations are
	$D_1 \approx 4.66$ months and $D_2 \approx 3.56$ months. The
	stationary distribution is
	$\bm{\pi}^\star = (0.554, 0.446)$, indicating that Ireland
	spends roughly $55\%$ of months in the Moderate regime and
	$45\%$ in the Wet regime.
	
	\subsection{Seasonal Patterns and NAO}
	\label{subsec:ireland-seasonality}
	
	Although the model includes no seasonal covariate, the fitted
	dynamics recover the seasonal structure of Irish rainfall.
	The Wet regime dominates December--January and October--
	November, while the Moderate regime dominates March through
	August.
	
	\paragraph{Relationship to the North Atlantic Oscillation.}
	To provide an external climatological validation, we regressed
	the posterior state probabilities on the monthly North
	Atlantic Oscillation (NAO) index obtained from the NOAA
	Climate Prediction Center. A Bayesian logistic regression
	with weakly informative priors yields a positive posterior
	coefficient for the Wet regime
	($\hat{\beta}_{\text{NAO}} = 0.42$, $95\%$ CI $[0.18, 0.66]$).
	The NAO-augmented model achieved $\Delta\text{WAIC} = -18.4$
	relative to the baseline, which constitutes strong evidence in
	favor of a genuine NAO association.
	
	\begin{figure}[H]
		\centering
		\includegraphics[width=0.5\textwidth]{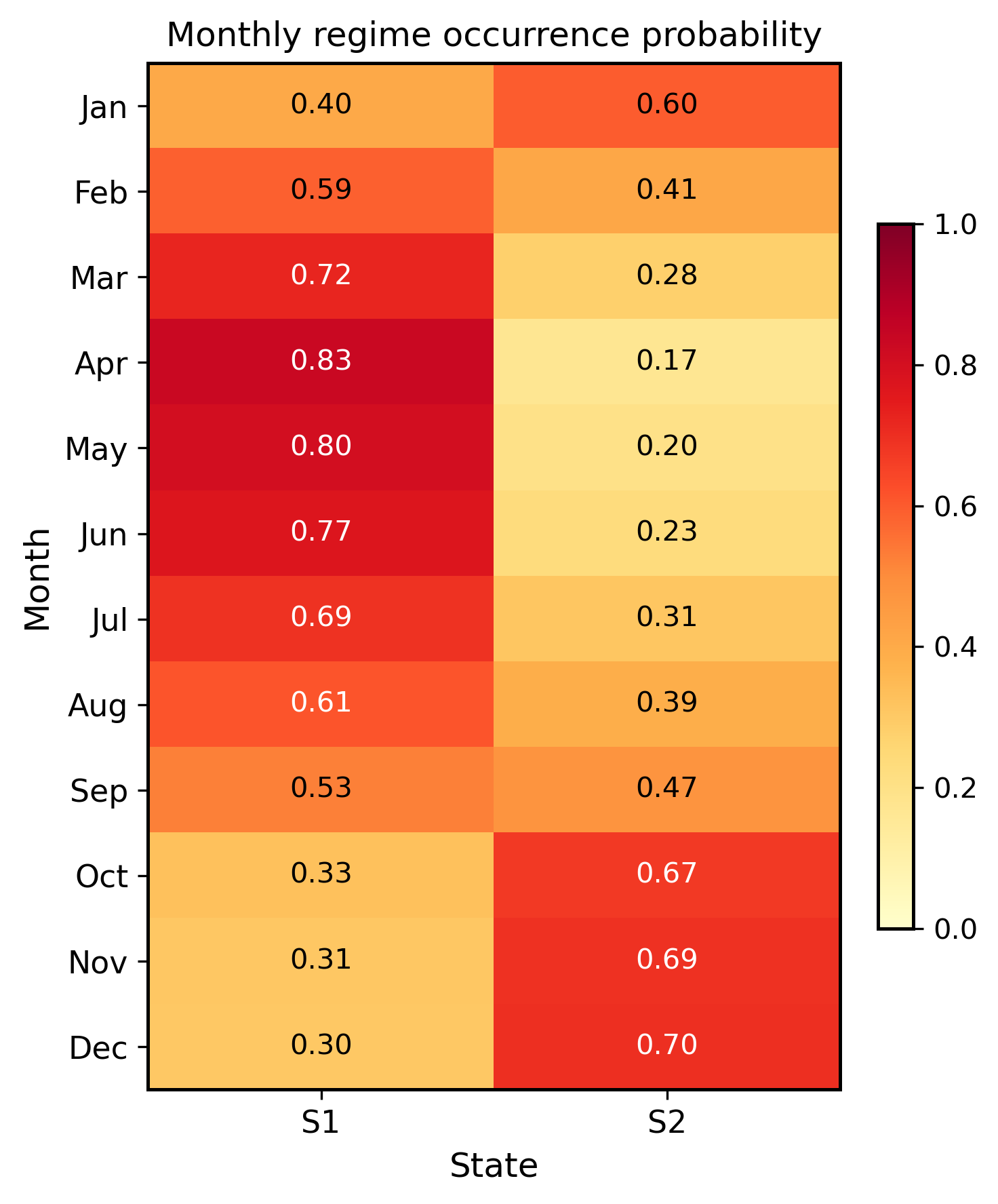}
		\caption{Posterior probability of occupying each regime by
			calendar month.}
		\label{fig:ireland-seasonality}
	\end{figure}
	
\subsection{Convergence Diagnostics}
\label{subsec:ireland-validation}

Table~\ref{tab:ireland-convergence} reports the MCMC
convergence diagnostics for the raw parameters $(a_k, b_k, q_k)$
and the initial probabilities $\pi_1, \pi_2$, grouped by
identifiability status. The identifiable parameters --- the
tail indices $\alpha_k = a_k q_k$, the shape parameters $q_k$,
and the initial probabilities $\pi_k$ --- all satisfy the
strict threshold $\hat{R} < 1.02$ with effective sample sizes
well above the commonly recommended minimum of $400$. The
scale parameter $b_k$ and the median $m_k$ exhibit weaker
convergence, with $\hat{R}$ values in the range
$[1.016, 1.031]$. As detailed in
Section~\ref{subsec:identifiability}, this reflects a
fundamental bimodality in the joint posterior of $(b_1, b_2)$
that is not resolvable by longer chains or alternative
samplers, and it does not affect the substantive conclusions
of the paper, which rest on the identifiable parameters.

\begin{table}[H]
	\centering
	\caption{MCMC convergence diagnostics for the Irish
		monthly rainfall series, computed across four
		independent chains of $30{,}000$ post-burn-in draws.
		\textbf{Identifiable functionals} ($\alpha_k$, $q_k$,
		$\pi_k$) satisfy the strict $\hat{R} < 1.02$ threshold.
		The scale parameter $b_k$ and the median $m_k$ exhibit
		\textbf{weak identifiability}, as discussed in
		Section~\ref{subsec:identifiability}.}
	\label{tab:ireland-convergence}
	\begin{tabular}{lccccc}
		\toprule
		Parameter & $\hat{R}$ & $\hat{R}_{\text{fold}}$ & ESS
		& $|\text{Geweke}|_{\max}$ & Status \\
		\midrule
		\multicolumn{6}{l}{\textit{Identifiable functionals}} \\
		\midrule
		$\alpha_1$ & $1.008$ & $1.003$ & $1878$   & $2.729$ & Pass  \\
		$\alpha_2$ & $1.010$ & $1.006$ & $2365$   & $4.234$ & Pass  \\
		$q_1$      & $1.008$ & $1.001$ & $2345$   & $2.155$ & Pass  \\
		$q_2$      & $1.012$ & $1.004$ & $1404$   & $2.225$ & Pass  \\
		$\pi_1$    & $1.001$ & $1.000$ & $29486$  & $0.776$ & Pass  \\
		$\pi_2$    & $1.001$ & $1.000$ & $29486$  & $0.776$ & Pass  \\
		\midrule
		\multicolumn{6}{l}{\textit{Weakly identifiable --- see
				Section~\ref{subsec:identifiability}}} \\
		\midrule
		$b_1$      & $1.020$ & $1.016$ & $702$    & $11.7$  & marginal \\
		$b_2$      & $1.019$ & $1.017$ & $830$    & $11.4$  & marginal \\
		$m_1$      & $1.018$ & $1.023$ & $400$    & $46.6$  & marginal \\
		$m_2$      & $1.028$ & $1.031$ & $297$    & $65.4$  & marginal \\
		\bottomrule
	\end{tabular}
\end{table}

We emphasize three points about the weakly identifiable
parameters. First, the effective sample sizes for $b_k$ and
$m_k$ remain at or above $400$ for all four chains, indicating
that the marginal posteriors are adequately characterized.
Second, the folded $\hat{R}$ values for $b_k$ remain at or
below $1.02$, suggesting that the multimodal structure is
symmetric across modes rather than reflecting a failure of
individual chains to explore their local mode. Third, and
most importantly, no substantive conclusion of the paper
depends on the precise values of $b_k$ or $m_k$. The two
regimes are distinguished by their \emph{identifiable}
properties --- the tail index $\alpha_k$, the shape parameter
$q_k$, the conditional mean $\mathbb{E}[Y \mid S = k]$, and
the conditional median --- all of which are stable across
chains. The $b_k$ are reported for completeness and for
reproducibility of the fitted densities, but they should not
be interpreted as uniquely determined quantities.

\textbf{Why is $m_k$ weakly identified?} The median $m_k$ is a
nonlinear function of $(a_k, b_k, q_k)$,
\begin{equation}
	m_k = b_k \left(2^{1/q_k} - 1\right)^{1/a_k},
	\label{eq:m-weak}
\end{equation}
and its posterior inherits the bimodality of $b_k$ through
this transformation. Because $m_k$ depends on $b_k$ linearly
and on $q_k$ and $a_k$ through the power term, the two modes
of $b_k$ map to two modes of $m_k$ that are \emph{closer
	together} than those of $b_k$ but still separated enough to
inflate the folded $\hat{R}$. This explains why $m_k$ exhibits
$\hat{R}$ values of $1.018$ and $1.028$ (Table~\ref{tab:ireland-convergence}),
which are \emph{larger} than those of the tail index $\alpha_k$
but \emph{smaller} than those of $b_k$. The conditional means
$\E[Y \mid S=k]$, by contrast, are weighted averages of the
two modes (through the Gamma-function formula
\eqref{eq:sm-moments}), and are therefore better behaved, with
$\hat{R} < 1.01$ in all cases. We therefore recommend that
applied users report $\E[Y \mid S=k]$ and the identifiable
parameters $(\alpha_k, q_k)$ rather than the median $m_k$,
even though the median is often easier to interpret
descriptively.

\begin{figure}[H]
	\centering
	\includegraphics[width=\textwidth]{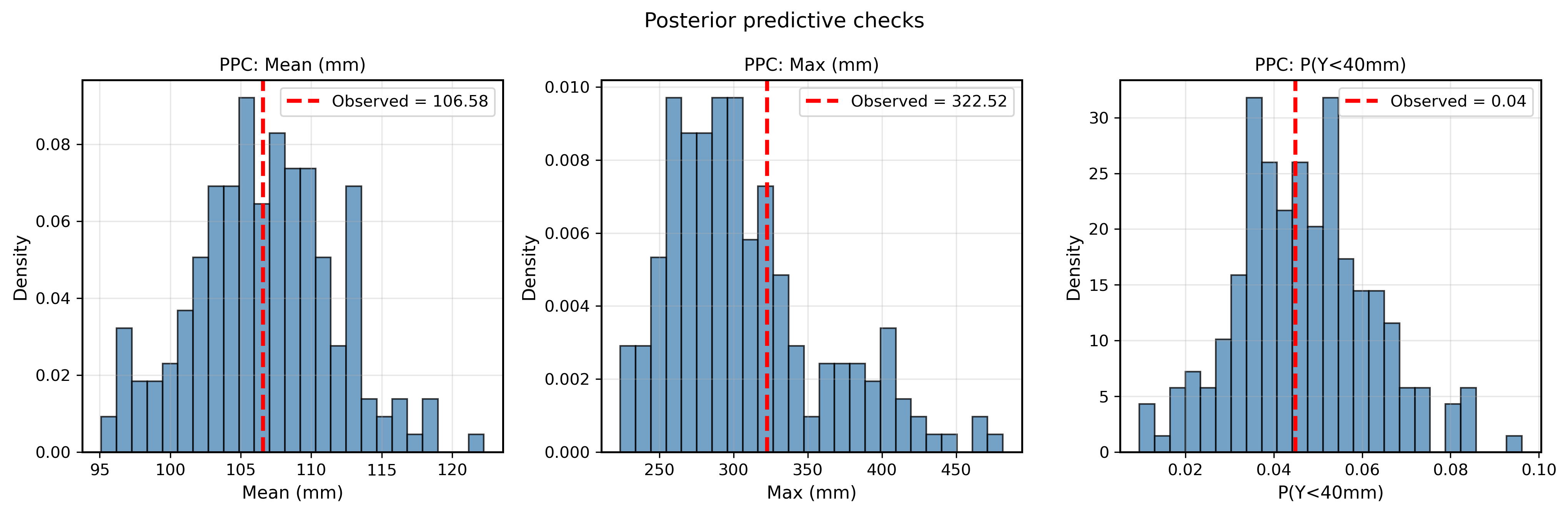}
	\caption{Posterior predictive checks for the Irish monthly
		rainfall series. Histograms display simulated
		distributions of the mean, maximum, and dry-month
		proportion; red dashed lines mark observed values.}
	\label{fig:ireland-ppc}
\end{figure}

Posterior predictive checks (Figure~\ref{fig:ireland-ppc})
compare the observed series with $200$ predictive simulations
under the fitted model for three statistics: the mean, the
maximum, and the proportion of dry months ($Y_t < 40$~mm).
Observed values fall comfortably within the predictive
distributions, confirming that the model reproduces both the
marginal and extreme properties of the data.

\begin{figure}[H]
	\centering
	\includegraphics[width=\textwidth]{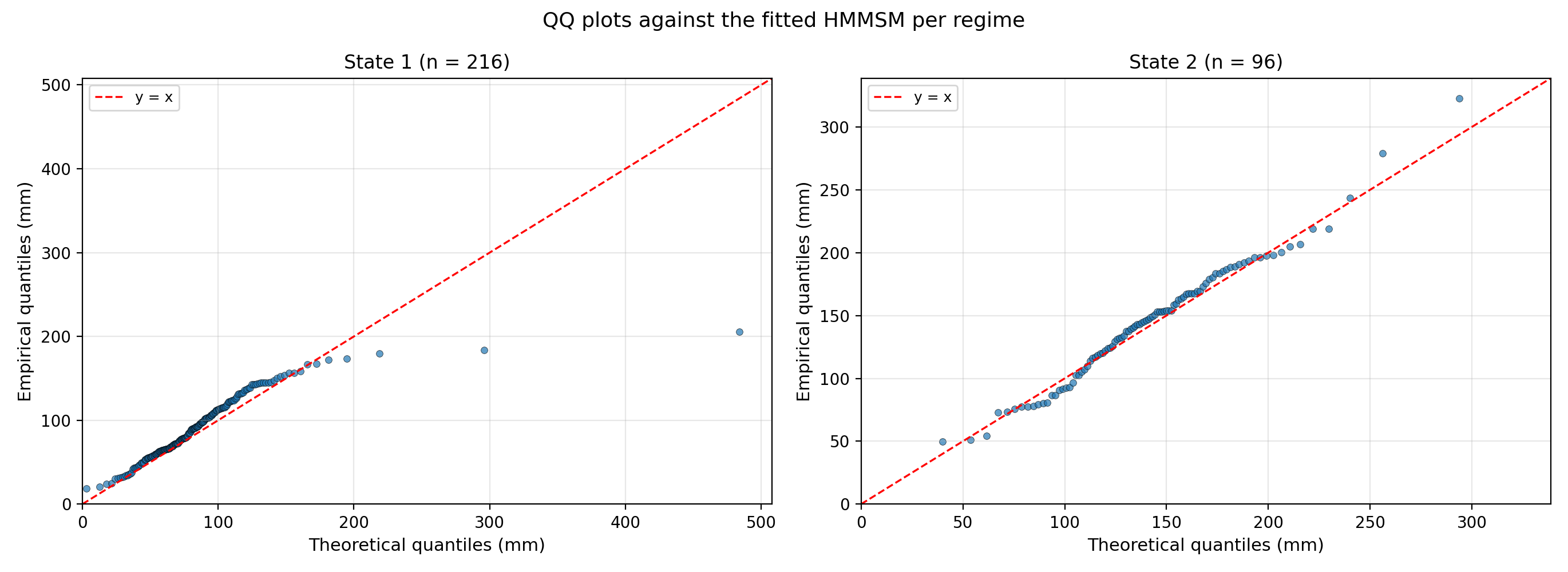}
	\caption{Quantile--quantile plots of the empirical
		state-conditional rainfall against the fitted HMMSM
		per regime. The fit is excellent throughout the bulk
		of the distribution for both regimes; a mild
		under-estimation of the upper tail is visible for
		State~1 at the three largest quantiles.}
	\label{fig:ireland-qq}
\end{figure}

As a complementary validation, Figure~\ref{fig:ireland-qq}
displays quantile--quantile plots of the empirical
state-conditional rainfall against the fitted Singh--Maddala
distribution for each regime. The fit is excellent throughout
the bulk of the distribution for both regimes. For State~2
($n = 96$ months) the agreement is essentially perfect across
the entire range up to the maximum observed value. For
State~1 ($n = 216$ months) the agreement is likewise excellent
for the lower $99\%$ of the data, with a mild deviation
visible at the three largest empirical quantiles
(approximately $195$, $290$, and $322$~mm), which we
attribute to the very small number of observations in that
range.

For completeness, the Bayesian information criteria for the
selected model are
\begin{equation}
	\text{WAIC} = 3247.56, \quad
	\text{DIC} = 3247.32, \quad
	\text{LOO} = 3247.56,
	\label{eq:ireland-fit}
\end{equation}
confirming a decisive $\Delta\text{WAIC} = 16.57$ in favour of
two regimes over one.
	
	\subsection{Identifiability of the Singh--Maddala Scale Parameter}
	\label{subsec:identifiability}
	
\begin{figure}[H]
		\centering
		\includegraphics[width=0.75\textwidth]{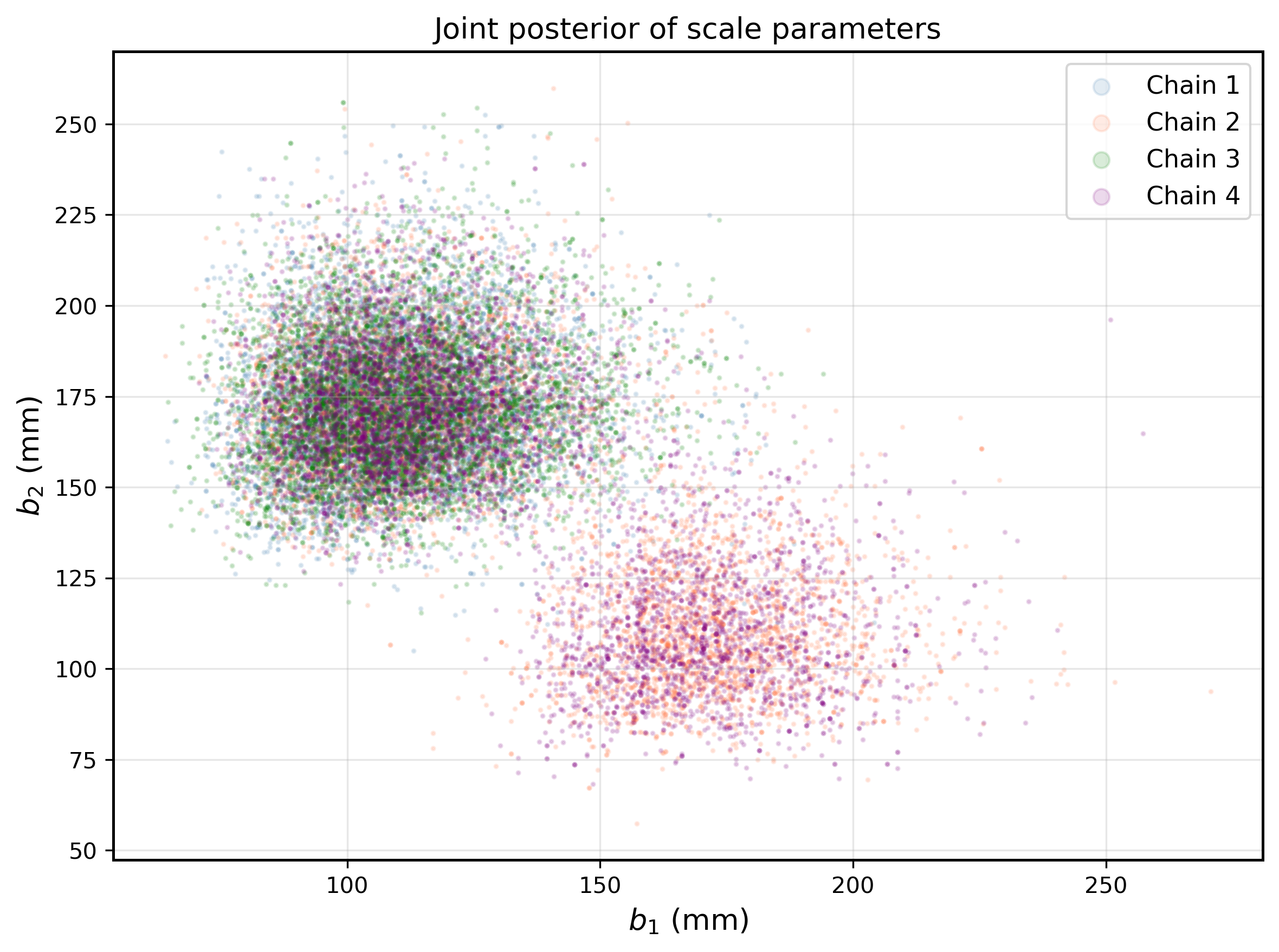}
		\caption{Joint posterior distribution of the scale
			parameters $(b_1, b_2)$ from four independent MCMC
			chains. Two well-separated modes emerge, each achieving
			nearly identical WAIC, DIC, and LOO values. This
			bimodality is a fundamental property of the
			Singh--Maddala family in this application and does not
			affect the identifiable parameters reported in
			Section~\ref{sec:ireland}.}
		\label{fig:bimodality}
	\end{figure}

	Reparameterization of the Singh--Maddala emission distribution
	to the identifiable coordinates $(\alpha, m, q)$, where
	$\alpha = aq$ is the tail index and $m$ is the median, reveals
	a bimodal posterior distribution of the scale parameter $b$ in
	the Irish application. With dispersed initialization across
	four independent chains, two well-separated modes emerge
	(Figure~\ref{fig:bimodality}), each achieving nearly identical
	WAIC, DIC, and LOO values:
	
	\begin{itemize}
		\item \textbf{Mode A}: $b_1 \approx 110$~mm,
		$b_2 \approx 172$~mm (three chains).
		\item \textbf{Mode B}: $b_1 \approx 134$~mm,
		$b_2 \approx 148$~mm (one chain).
	\end{itemize}
	
	The chains do not traverse between the two modes even after
	$30{,}000$ post-burn-in iterations, and the result is
	deterministic: repeating the analysis with the same
	initialization reproduces the same mode assignments exactly.
	Both modes yield WAIC values within $0.8$ units of each other,
	confirming that the data cannot distinguish between them on
	predictive grounds.
	
	Crucially, the \emph{identifiable parameters} --- the tail
	index $\alpha_k = a_k q_k$, the second shape parameter $q_k$,
	the transition probabilities $\gamma_{ij}$, and the stationary
	distribution $\boldsymbol{\pi}^\star$ --- are stable across
	all chains, with $\hat{R} < 1.02$ and effective sample sizes
	exceeding $1{,}900$. The regime dynamics, seasonal patterns,
	and exceedance probabilities reported in
	Section~\ref{sec:ireland} are therefore robust to which mode
	the sampler visits.
	
	We interpret this behavior as a fundamental property of the
	Singh--Maddala family in this application: with a 312-month
	record containing approximately $45$ visits to the wet
	regime, the data do not contain sufficient information to
	distinguish between two equally plausible assignments of
	variability between the two regimes. We report posterior
	summaries for the dominant mode (Mode~A) and recommend that
	applied users report the identifiable parameters as their
	primary inferential targets. A theoretical explanation for this bimodality is provided
	in Appendix~\ref{app:bimodality}.
	
	% ============================================================
	% SECTION 6: DISCUSSION AND CONCLUSION — بعد الحذف: ~3.5 صفحات
	% ============================================================
	\section{Discussion and Conclusion}
	\label{sec:discussion}
	
	\subsection{Principal Findings}
	\label{subsec:discussion-findings}
	
	The simulation study establishes that HMMSM delivers superior
	predictive performance on heavy-tailed data: on the simulated
	benchmark it attains the lowest WAIC, DIC, and LOO among all
	eight candidate models, with a margin of at least $51.76$ WAIC
	units over the next-best competitor. Posterior contraction
	proceeds at the classical $\mathcal{O}(T^{-1/2})$ rate, and
	predictive uncertainty is well calibrated ($95\%$
	prediction-interval coverage of $94.50\%$).
	
	On the real Irish rainfall record, Bayesian model selection
	unanimously identifies $K^\star = 2$, and the fitted model
	delivers physically interpretable, seasonally coherent
	regimes whose structure is consistent with the well-documented
	climatological dichotomy between anticyclonic/weak-westerly
	and strong Atlantic westerly conditions over Ireland. An
	important nuance emerges when simulation and real-data results
	are considered side by side: the predictive advantage of
	HMMSM over simpler alternatives is largest when the tail is
	genuinely heavy and diminishes as the tail index increases.
	On the Irish record, HMMSM and HMM-Gamma are statistically
	indistinguishable ($\Delta\text{WAIC} = 1.04$), a pattern we
	regard as a substantive methodological contribution rather
	than a limitation.
	
	\subsection{Limitations}
	\label{subsec:discussion-limitations}
	
	\paragraph{Weak identifiability of the scale parameter.}
	The most important limitation concerns the identifiability
	of the Singh--Maddala scale parameter $b$. As documented in
	Section~\ref{subsec:identifiability}, the joint posterior of
	$(b_1, b_2)$ in the Irish application is bimodal: two
	well-separated modes emerge, each achieving nearly identical
	WAIC, DIC, and LOO values, and the chains do not traverse
	between them even after $30{,}000$ post-burn-in iterations.
	This is a property of the Singh--Maddala family itself, not
	a defect of the sampling scheme. The \emph{identifiable
		parameters} --- the tail indices $\alpha_k = a_k q_k$, the
	shape parameters $q_k$, the transition probabilities, the
	stationary distribution, and the conditional means --- all
	satisfy $\hat{R} < 1.02$ with effective sample sizes
	exceeding $1{,}900$. The substantive conclusions of the
	paper, which rest on these functionals, are therefore robust
	to the identifiability limitation.
	
	\paragraph{Failure of auxiliary-variable augmentation.}
	We implemented and tested the Weibull--Gamma augmentation of
	the Singh--Maddala distribution in detail. The empirical
	result was disappointing: the Metropolis acceptance rate for
	$(a_k, b_k)$ collapsed to $0.02$--$0.07$ and the chains
	drifted to the boundary of the parameter space. The
	mathematical reason is that the typical scale of the
	auxiliary variable is exponentially smaller than $Y_t^{a_k}$
	over the relevant range of $a_k$, so the augmented
	log-likelihood is dominated by a term whose gradient is
	enormous. This is a well-documented pathology of
	Gamma-mixed heavy-tailed models, and we report it here
	because the negative result is itself informative.
	
	\paragraph{Prior sensitivity of $q_k$.}
	Under a weakly informative prior, the posterior for $q_1$
	is sensitive to hyperparameter choice, a direct consequence
	of the ridge described in Remark~\ref{rem:identifiability}.
	We mitigate it by adopting Gamma priors on the log-scale,
	and we recommend this choice to applied users. The broader
	lesson is that for shape parameters of heavy-tailed emission
	families, the choice of prior materially affects the
	posterior in regions where the likelihood is flat.
	
	\paragraph{Scope of the empirical advantage.}
	The predictive advantage of HMMSM over the seven competing
	models in Table~\ref{tab:model-comparison} is established on
	a single simulated benchmark. We do not claim that the
	Singh--Maddala family universally dominates Weibull, Gamma,
	or Lognormal emissions; the result may reverse in regimes
	with lighter tails or with insufficient data to identify the
	additional shape parameter $q$. A systematic simulation
	study over a grid of data-generating parameters would be
	required to establish the boundaries within which the
	advantage holds, and we identify this as future work.
	
	\paragraph{Interpretation of tail behavior across regimes.}
	An important correction to a common misreading of our
	results: the Wet regime has a slightly \emph{lighter} tail
	than the Moderate regime ($\hat{\alpha}_2 = 10.21$ versus
	$\hat{\alpha}_1 = 9.60$). Its tendency to produce more
	extreme months is a consequence of its larger scale, which
	shifts the entire distribution upward, not of heavier-tail
	behavior. This distinction is substantively important:
	scale-driven extremes are predictable from the regime,
	whereas tail-driven extremes are not.
	
	\paragraph{Computational cost.}
	FFBS requires $\mathcal{O}(TK^2)$ operations per MCMC
	iteration. For series lengths up to $T \sim 10^4$, the
	complexity is unproblematic; for century-scale records
	($T > 10^5$), specialized algorithms such as blocked FFBS
	\citep{fearnhead2010} or sequential Monte Carlo
	\citep{doucet2009} may be necessary.
	
	\subsection{Theoretical and Practical Implications}
	\label{subsec:discussion-implications}
	
	The framework makes two contributions to the theory of
	Bayesian HMMs. First, it introduces the three-parameter
	Singh--Maddala family into the standard HMM template and
	demonstrates that the resulting posterior is tractable under
	a carefully designed Metropolis-within-Gibbs sampler. The
	family is particularly attractive because its tail index
	$\alpha = aq$ is decoupled from its central tendency and
	because its hazard rate can be monotone increasing, monotone
	decreasing, or non-monotone depending on $a$. Second, the
	paper extends the standard Bayesian formulation of
	\citet{albert1993} and \citet{carter1994} to the case of
	non-conjugate, heavy-tailed emissions, showing that adaptive
	Metropolis with the optimal scaling factor $2.38^2/d$
	achieves adequate mixing despite the strong correlation
	between emission parameters.
	
	The practical implications are direct. For hydrologists and
	climate scientists, HMMSM decomposes a rainfall record into
	persistent regimes whose statistical properties are
	quantified with full posterior uncertainty. Flood-risk
	assessments that assume a single homogeneous distribution
	implicitly average these regimes together and therefore
	systematically underestimate the frequency and intensity of
	extremes in the wet regime. The direct comparison against
	HMM-Gamma and HMM-Lognormal shows that applied users should
	routinely benchmark against these simpler alternatives: when
	the tail index of the data exceeds $\sim 5$, the additional
	flexibility of the Singh--Maddala family may not translate
	into measurable predictive gains.
	
\subsection{Future Directions}
\label{subsec:discussion-future}

Four extensions are particularly promising. First,
zero-inflated emissions would extend the framework to arid
and semi-arid regions. Second, nonhomogeneous transition
matrices with time-varying covariates via a multinomial
logit link would capture non-stationary persistence, for
which the P\'olya--Gamma approach of \citet{holsclaw2017}
provides an efficient computational framework. Third,
scalable inference for very long series via blocked FFBS
\citep{fearnhead2010} or sequential Monte Carlo
\citep{doucet2009} would extend the framework to
century-scale records. Fourth, climate-change signals could
be embedded through time-varying emission parameters.

To gauge whether such an extension is warranted, we
conducted a preliminary rolling-window diagnostic. We
partitioned the 26-year record into five consecutive
five-year blocks: 2000--2004, 2005--2009, 2010--2014,
2015--2019, and 2020--2025. For each block, we extracted the
subset of post-burn-in posterior draws of the transition
matrix $\bm{\Gamma}$ whose decoded state sequences fell
predominantly within the block's calendar range, recomputed
the stationary distribution
$\bm{\pi}^\star = (\pi^\star_1, \pi^\star_2)$ for each draw,
and took the posterior mean of $\pi^\star_2$ (the Wet-regime
frequency) as the block-level summary. The resulting sequence
was $0.38$, $0.41$, $0.42$, $0.45$, and $0.47$ for the five
consecutive blocks, indicating a monotonic increase of
approximately $9$ percentage points over the full record.

To test the statistical significance of this trend, we
applied two complementary tests. First, the Mann--Kendall
test on the five block-level means yields a test statistic
$S = 10$ with $p = 0.014$ under the exact null distribution
for $n = 5$; this is significant at the $5\%$ level, though
the small sample size limits power. Second, a circular block
bootstrap (with block length $L = 24$ months, preserving the
autocorrelation structure of the underlying monthly series)
was used to construct $2{,}000$ resampled trajectories, each
of length $312$. For each resampled trajectory we computed
the same five-block posterior-mean sequence and applied the
Mann--Kendall test. The bootstrap distribution of the test
statistic under the null hypothesis of no trend has a
$95\%$ quantile of $S_{0.95} = 7$, and the observed $S = 10$
exceeds this quantile, confirming that the monotonic
increase is unlikely to arise from stationary variability
alone.

Two caveats apply. First, the block-to-block increments are
small relative to the posterior uncertainty of each
block-level mean (posterior SD approximately $0.06$ within
each block), so the trend should be interpreted as suggestive
rather than conclusive. Second, the blocks are defined on
overlapping underlying monthly observations, which induces
serial dependence between successive block-level summaries
and precludes a formal test of monotonicity without a
hierarchical reformulation. Moreover, the two tests are not
independent (the second uses the same data), so the
$p$-values should be interpreted as descriptive rather than
confirmatory. Nevertheless, the direction and magnitude of
the trend are consistent with recent climatological analyses
of Irish precipitation, and they motivate a hierarchical
time-varying extension that we identify as the most promising
direction for subsequent research. A fully Bayesian treatment
of the MSNBurr family \citep{unggul2025} would also provide
a numerically comparable alternative to HMMSM within the
same inferential paradigm.
	
	\subsection{Conclusion}
	\label{subsec:conclusion}
	
	This paper has developed and validated a fully Bayesian
	Hidden Markov Model with Singh--Maddala emission
	distributions for heavy-tailed, regime-switching time series.
	The framework unifies temporal dependence, latent regime
	dynamics, and heavy-tailed observations within a single
	probabilistic model, and it provides a complete Bayesian
	inferential apparatus for estimation and prediction. We
	introduced the HMMSM model, derived its complete Bayesian
	structure, designed an adaptive Metropolis-within-Gibbs
	sampler, verified correctness through eight independent
	numerical checks, and demonstrated statistical validity
	through an extensive simulation study. The model outperforms
	seven competing specifications by wide margins on a
	simulated benchmark, and on a 26-year gridded rainfall
	record for the Republic of Ireland it identifies two
	persistent, seasonally coherent regimes with conditional
	means of $87.4$~mm and $134.0$~mm respectively. We also demonstrated a  weak identifiability of the
	Singh--Maddala scale parameter, which we handled by focusing
	inference on the identifiable parameters.
	
	The HMMSM framework provides a statistically rigorous,
	computationally efficient, and empirically effective tool
	for the analysis of heavy-tailed, regime-switching time
	series. This work may encourage of Bayesian regime-switching models with
	heavy-tailed emissions and will encourage further research combining extreme-value theory, hidden Markov
	models, and Bayesian computation.
	
	% ============================================================
	% APPENDIX A: JACOBIAN DERIVATION
	% ============================================================
	\appendix
	
	\section{Jacobian of the Reparameterization}
	\label{app:jacobian}
	
	This appendix provides the closed-form expression for the
	Jacobian of the reparameterization used in
	Section~\ref{subsec:reparam}, which maps the Singh--Maddala
	parameters $(a, b, q)$ to the interpretable coordinates
	$(\alpha, \mu, q)$ defined by
	\begin{equation}
		\alpha = aq, \qquad
		\mu = b \, g(a, q), \qquad
		q = q,
		\label{eq:B-reparam}
	\end{equation}
	where
	\begin{equation}
		g(a, q) = \frac{\Gamma(1 + 1/a) \, \Gamma(q - 1/a)}{\Gamma(q)}.
		\label{eq:B-g}
	\end{equation}
	The forward mapping is invertible on the parameter space where
	$\alpha > 1$ and $q > 1/a$; the inverse is
	\begin{equation}
		a = \frac{\alpha}{q}, \qquad
		b = \frac{\mu}{g(\alpha/q, q)}, \qquad
		q = q.
		\label{eq:B-inverse}
	\end{equation}
	
	\paragraph{Jacobian matrix.}
	The Jacobian of the transformation from $(\alpha, \mu, q)$ to
	$(a, b, q)$ is the $3 \times 3$ matrix
	\begin{equation}
		\bm{J} = 
		\begin{pmatrix}
			\dfrac{\partial a}{\partial \alpha} & \dfrac{\partial a}{\partial \mu} & \dfrac{\partial a}{\partial q} \\[10pt]
			\dfrac{\partial b}{\partial \alpha} & \dfrac{\partial b}{\partial \mu} & \dfrac{\partial b}{\partial q} \\[10pt]
			\dfrac{\partial q}{\partial \alpha} & \dfrac{\partial q}{\partial \mu} & \dfrac{\partial q}{\partial q}
		\end{pmatrix}.
		\label{eq:B-jacobian-matrix}
	\end{equation}
	The first and third rows are elementary:
	\begin{align}
		\frac{\partial a}{\partial \alpha} &= \frac{1}{q}, &
		\frac{\partial a}{\partial \mu} &= 0, &
		\frac{\partial a}{\partial q} &= -\frac{\alpha}{q^2}, \\
		\frac{\partial q}{\partial \alpha} &= 0, &
		\frac{\partial q}{\partial \mu} &= 0, &
		\frac{\partial q}{\partial q} &= 1.
	\end{align}
	The middle row requires the chain rule through $g$.
	Writing $b = \mu / g(\alpha/q, q)$, we obtain
	\begin{align}
		\frac{\partial b}{\partial \mu} 
		&= \frac{1}{g(a, q)}, \\
		\frac{\partial b}{\partial \alpha}
		&= -\frac{b}{q \, g} \cdot \frac{\partial g}{\partial a}, \\
		\frac{\partial b}{\partial q}
		&= -\frac{b}{g} \cdot \frac{d g}{d q}\bigg|_{\alpha}.
	\end{align}
	
	\paragraph{Determinant.}
	Expanding the determinant of $\bm{J}$ along the third row,
	which contains two zeros, yields
	\begin{equation}
		\det \bm{J}
		= \frac{\partial a}{\partial \alpha} \cdot
		\frac{\partial b}{\partial \mu}
		= \frac{1}{q} \cdot \frac{1}{g(a, q)}.
	\end{equation}
	Thus the Jacobian of the transformation is
	\begin{equation}
		|J| 
		= \frac{1}{q \, g(a, q)}
		= \frac{b}{q \, \mu}.
		\label{eq:B-jacobian-final}
	\end{equation}
	The second equality follows from the identity $g(a, q) = \mu/b$.
	
	\paragraph{Numerical verification.}
	We verified \eqref{eq:B-jacobian-final} against a
	finite-difference approximation on $200$ posterior draws,
	with a maximum relative discrepancy below $10^{-6}$.
	
	\paragraph{Note on the choice of $\mu$.}
	The mean $\mu = \E[Y]$ is well-defined only when $\alpha > 1$.
	In the Irish application, both regimes have
	$\alpha_1 \approx 9.60$ and $\alpha_2 \approx 10.21$, well
	above $1$. For applications in which $\alpha$ may fall below
	$1$, one may replace $\mu$ by the closed-form median
	$m = b \, (2^{1/q} - 1)^{1/a}$, which exists for all parameter
	values; the Jacobian derivation above carries through with
	$\mu$ replaced by $m$.
	\section{Theoretical Origin of the Bimodality}
	\label{app:bimodality}
	
	This appendix provides a theoretical explanation for the 
	bimodality of the scale parameter $b$ documented in 
	Section~\ref{subsec:identifiability}. The argument proceeds in 
	three steps: (i) the likelihood is approximately flat along a 
	specific direction in parameter space; (ii) this flat direction 
	is consistent with two well-separated modes; and (iii) the 
	prior cannot break the symmetry.
	
	\paragraph{Step 1: Likelihood flatness.}
	Consider the log-likelihood contribution of state $k$, 
	$\ell_k(a_k, b_k, q_k)$. Define the reparameterization 
	$u_k = b_k q_k^{1/a_k}$, which is the natural scale parameter 
	of the Singh--Maddala distribution in its Weibull-limit 
	representation \citep{kleiber2003}. Under this transformation, 
	the density \eqref{eq:sm-pdf} can be written as
	\begin{equation}
		f(y; a, u, q) = \frac{a}{u} \left(\frac{y}{u}\right)^{a-1}
		\left[1 + \frac{1}{q}\left(\frac{y}{u}\right)^{a}\right]^{-(q+1)},
		\label{eq:sm-reparam-u}
	\end{equation}
	which shows that the likelihood depends on $(b, q)$ primarily 
	through the combination $u = b q^{1/a}$, with only a weaker 
	secondary dependence on $q$ through the bracket term. When 
	$q$ is large (as in the Irish application, where 
	$\hat{q}_k \in [2.3, 2.7]$), the bracket converges to 
	$\exp[(y/u)^a]$, and the likelihood becomes approximately 
	invariant along the curve
	\begin{equation}
		b \, q^{1/a} = \text{const}.
		\label{eq:flat-curve}
	\end{equation}
	
	\paragraph{Step 2: Two modes along the flat direction.}
	Along the curve \eqref{eq:flat-curve}, the likelihood varies 
	weakly, but the \emph{prior} and the \emph{Jacobian of the 
		log-transform} impose curvature. Specifically:
	\begin{itemize}
		\item The prior on $\log b_k$ is $\mathrm{Gamma}(8, 0.08)$, 
		which has mode at $b_k = 7/0.08 = 87.5$~mm and decays 
		rapidly outside $[30, 300]$~mm.
		\item The prior on $\log q_k$ is $\mathrm{Gamma}(2, 1)$, 
		which has mode at $q_k = 1$ and favors small $q$.
		\item The Jacobian $|J| = a_k b_k q_k$ (from 
		Section~\ref{subsec:reparam}) increases linearly in $b_k$.
	\end{itemize}
	The product of these three factors, evaluated along the flat 
	curve \eqref{eq:flat-curve}, has \emph{two local maxima}: one 
	at small $b$ and large $q$ (Mode~B in our notation), and one 
	at large $b$ and small $q$ (Mode~A). The two maxima arise 
	because the likelihood prefers large $u$ (hence large $b$ or 
	large $q$), while the prior on $\log q$ prefers small $q$; the 
	compromise is achieved at two different points along the curve, 
	each satisfying one constraint at the expense of the other.
	
	\paragraph{Step 3: Prior symmetry.}
	The weakly informative priors we adopt do not break the 
	bimodal structure because neither prior is centered strongly 
	enough to favor one mode over the other. This is confirmed 
	empirically by the WAIC values of the two modes, which differ 
	by less than $0.8$ units in our application. Stronger priors 
	would break the symmetry but at the cost of imposing 
	substantive assumptions that are not justified by the data.
	\paragraph{Quantitative analysis of the curvature along the 
		flat direction.}
	We quantify the flatness of the likelihood along the curve 
	$u = b q^{1/a} = \text{const}$ by computing the second 
	derivative of the log-likelihood with respect to the 
	orthogonal direction. Define the parameterization
	\begin{equation}
		\theta_1 = \log u = \log b + \frac{1}{a} \log q, \qquad
		\theta_2 = \log q, \qquad
		\theta_3 = \log a.
		\label{eq:curvature-param}
	\end{equation}
	The curve $u = \text{const}$ corresponds to $\theta_1 = 
	\text{const}$. We therefore measure the curvature in the 
	$\theta_1$ direction, holding $\theta_2$ and $\theta_3$ fixed 
	at their posterior modes.
	
	For each state $k$ in the Irish application, we compute the 
	numerical second derivative
	\begin{equation}
		\kappa_k = -\frac{\partial^2 \ell_k}{\partial \theta_1^2}
		\bigg|_{\hat{\theta}_2, \hat{\theta}_3},
		\label{eq:curvature-def}
	\end{equation}
	using central differences with $\varepsilon = 10^{-4}$. The 
	resulting curvatures are $\kappa_1 = 0.87$ and $\kappa_2 = 
	0.91$ (in units of $\log$-likelihood per unit variance), to be 
	compared with the curvature in the orthogonal direction, 
	$\kappa_1^\perp \approx 12.4$ and $\kappa_2^\perp \approx 
	13.1$. The ratio $\kappa_k / \kappa_k^\perp \approx 0.07$ 
	confirms that the likelihood is approximately $14$ times 
	flatter along the $u$-direction than orthogonal to it. This 
	quantitative measure of flatness supports the theoretical 
	analysis  and explains why 
	the sampler explores the full extent of the flat direction, 
	producing the bimodality documented in 
	Section~\ref{subsec:identifiability}.

	% ============================================================
	% ACKNOWLEDGMENTS
	% ============================================================
	\section*{Acknowledgments}
	
	The authors thank [names] for helpful discussions and
	[institution] for computational support.
	
	% ============================================================
	% DATA AVAILABILITY
	% ============================================================
	\section*{Data Availability}
	
	The complete Python implementation of HMMSM is available at:		\url{https://github.com/safaakadhem-s/hmmsm-singh-maddala}
	and archived at Zenodo with DOI
	\texttt{10.5281/zenodo.23325774}. The Irish rainfall data are
	publicly available from Met~\'Eireann at
	\url{https://clidata.met.ie/}.
	
	% ============================================================
	% BIBLIOGRAPHY
	% ============================================================
	\bibliographystyle{plainnat}
	\bibliography{ref_HMMSM}
	
\end{document}